\documentclass[prd,a4paper,showpacs,
nofootinbib
]{revtex4}
\usepackage{graphicx}
\usepackage{epsfig}

\newcommand{\beq}{\begin{equation}}
\newcommand{\eeq}{\end{equation}}
\newcommand{\bea}{\begin{eqnarray}}
\newcommand{\eea}{\end{eqnarray}}
\newcommand{\etal}{{\it et al.}}
\def\t{{\tan^2\theta_{13}}}
\def\lsim{\ \raisebox{-.4ex}{\rlap{$\sim$}} \raisebox{.4ex}{$<$}\ }
\def\gsim{\ \raisebox{-.4ex}{\rlap{$\sim$}} \raisebox{.4ex}{$>$}\ }
\def\kl{{KamLAND~}}
\def\nue{{\nu_e}}
\def\anue{{\bar\nu_e}}
\def\numu{{\nu_{\mu}}}
\def\anumu{{\bar\nu_{\mu}}}
\def\nutau{{\nu_{\tau}}}
\def\anutau{{\bar\nu_{\tau}}}
\def\p2eboplus{\bar{P}^\oplus_{2e}}
\def\p1eboplus{\bar{P}^\oplus_{1e}}

\newcommand{\peenh}{\mbox{$P_{ee}^{NH}$~}}
\newcommand{\peebarnh}{\mbox{P$_{\bar{e}\bar{e}}^{NH}$}}
\newcommand{\peeih}{\mbox{$P_{ee}^{IH}$~}}
\newcommand{\peebarih}{\mbox{P$_{\bar{e}\bar{e}}^{IH}$}}
\newcommand{\ptoe}{\mbox{$P^\oplus_{2e}$}}
\newcommand{\ponee}{\mbox{$P^\oplus_{1e}$}}

\newcommand{\sabsq}{\mbox{$\sin^{2}\theta_{12}~$}}
\newcommand{\poneeb}{\mbox{$\bar{P}^\oplus_{1e}$}}

\newcommand{\cacsq}{\mbox{$\cos^{2}\theta_{13}~$}}

\begin{document}

\begin{flushright}
SINP/TNP/03-44/\\
SISSA 114/2003/EP\\
hep-ph/0312315
\end{flushright}

\title{Prospects of probing $\theta_{13}$ and neutrino mass hierarchy by  
Supernova Neutrinos in KamLAND} 

\author{Abhijit Bandyopadhyay$^1$}
\author{Sandhya Choubey$^{2,3}$}
\author{Srubabati Goswami$^4$}
\author{Kamales Kar$^{1,5}$}
\affiliation{$^1$Theory Group, Saha Institute of Nuclear Physics,
1/AF, Bidhannagar,
Calcutta 700 064, INDIA}
\affiliation{$^2$INFN, Sezione di Trieste, Trieste, Italy} 
\affiliation{$^3$Scuola Internazionale Superiore di Studi Avanzati,
I-34014, Trieste, Italy}
\affiliation{$^4$Harish-Chandra Research Institute, Chhatnag Road, Jhusi,
Allahabad  211 019, INDIA}
\affiliation{$^5$The Institute of Mathematical Sciences, C.I.T. Campus,
Taramani,
Chennai 600113, INDIA}


\begin{abstract} 
\vspace {10pt}
In this paper we study the physics potential of the \kl detector 
in probing neutrino oscillation parameters through observation of supernova 
neutrinos. In particular, we discuss the possibilities of probing 
the mixing angle $\theta_{13}$ and 
determining the sign of $\Delta m^2_{32}$
from the total charged current(CC) event rates on the proton and 
$^{12}{C}$ target, as well as from the CC spectra. 
We discuss the chances of probing the earth matter 
effect induced modulations from the observation of 
CC spectra in the 
different CC reactions in \kl and find the volume required 
to get a statistically significant signature of the earth matter 
effect in different energy bins. 
We also calculate the event rates expected in the 
neutral current (NC) reactions on Carbon and free proton 
and investigate if the charged current to neutral current ratios, 
which are free of the absolute luminosity uncertainty in the supernova 
neutrino fluxes, 
can be useful in probing the oscillation parameters. 

\end{abstract}

\maketitle

\section{Introduction}
\kl is a 1 kton liquid scintillator detector situated in the 
Kamioka mine in Japan. This detector is at present engaged in 
measuring the flux of electron antineutrinos 
coming from the various nuclear reactors in Japan. 
In 2002 it reported a phenomenal 
result
\cite{kl},
providing the clinching evidence for 
the Large-Mixing-Angle(LMA) MSW \cite{msw} solution
to the solar neutrino problem 
-- which was emerging as the favoured solution
to explain the solar neutrino shortfall 
in Homestake, 
Kamiokande, SAGE, GALLEX/GNO \cite{sol}, Super-Kamiokande(SK)
\cite{SKsolar} and the Sudbury Neutrino Observatory(SNO) 
\cite{sno1,sno2,sno3} experiments. 
The combined two-neutrino oscillation
analyses of the
solar neutrino and KamLAND  data including the most 
recent SNO salt results give the best-fit values of neutrino 
mass squared  difference  and  
mixing angle as 
$\Delta m^2_{\odot}(\equiv \Delta m^2_{21})
=7.2 \times 10^{-5}~{\rm eV^2}$, 
$\sin^2\theta_\odot (\equiv \sin^2\theta_{12}) = 0.3$ \cite{sno3,snosalt}.
The neutral current data from SNO had been instrumental in ruling out the 
maximal mixing and dark-side solutions ($\theta_{\odot} \geq \pi/4$),
implying $\Delta m^2_{21} > 0$  
\cite{sno2,snoncus},
a result now confirmed
at more than $5\sigma$ level by the latest results from SNO
\cite{sno3,snosalt}. 
 
Compelling evidence in favour of neutrino oscillation 
have also come from the 
observation of up-down asymmetry in the Zenith angle 
distribution of atmospheric neutrinos in the 
SK detector.
The best description to the data comes in terms of 
$\nu_{\mu} \rightarrow \nu_{\tau}$
($\bar{\nu}_{\mu}\rightarrow \bar{\nu}_{\tau}$) oscillations
with  maximal mixing and $\Delta m^2_{atm}$($\equiv \Delta m^2_{32}$)
$=2.0 \times 10^{-3}$ eV$^2$ 
from a recent re-analysis of data performed by the 
SK collaboration \cite{SKatmo03}.

In the realistic  three neutrino oscillation scenario
the two sectors -- solar and atmospheric -- 
get coupled by the mixing angle $\theta_{13}$, which
is at present constrained by the reactor data
as $\sin^2 \theta_{13} \leq 0.03$ at 90\% C.L. \cite{chooz}.
This is only an upper bound and one of the major issues to be settled 
by future experiments is -- how small is $\theta_{13}$.
Another important question which has still not 
been ascertained by the current
data is the sign of $\Delta m^2_{31}$.
Both these issues are expected to be addressed in future long baseline 
experiments 
\cite{Itow:2001ee,Ayres:2002nm}. 
Reactor experiments have also been shown
recently to have promising  capabilities to probe $\theta_{13}$
\cite{Minakata:2002jv,Huber:2003pm} 
and under very specific demanding conditions 
also the sign of $\Delta m^2_{32}$ \cite{piai}. 
A core collapse supernova (SN) in our galaxy can also  
play a significant role in throwing light on the above two issues 
\cite{Dighe:1999bi,Takahashi:2001ep,Lunardini:2003eh}.
The  feasibility of supernova neutrino detection has already been 
demonstrated by the 
chance recording of the 
neutrino events from the supernova SN1987A 
by Kamiokande \cite{Hirata:1987hu} and IMB 
\cite{Bionta:1987qt}
detectors.
The few neutrino events detected 
were sufficient to confirm the 
basic theory of core collapse supernova and at the same time could be used to 
set limits on neutrino properties. 
SN1987A  occurred in the Large Magellanic cloud at a distance of about 50 
kilo parsec(kpc) from earth. 
For a supernova in our galactic center,
we can assume a typical distance of
10 kpc and therefore expect many more neutrino events.
In type II supernovae the explosion takes a
few hours to penetrate through the envelope and only after that 
the optical
and other electromagnetic signals emerge. Whereas neutrinos start coming
out right after core bounce 
and are therefore useful for alerting supernova telescopes.
Therefore it is important that the present detectors be ready 
and the potentials fully explored before the next galactic supernova 
occurs. 
The capabilities of present and future detectors 
towards detection of neutrinos from a  galactic supernova 
has been investigated 
by many authors (see e.g. \cite{Sharp:2002as} for a list of references). 
In particular, 
the ability of SN neutrinos to probe oscillation parameters  
have been considered in many papers 
\cite{Dighe:1999bi,Takahashi:2001ep,Lunardini:2003eh,snpapers}. 
For example, it is shown in \cite{Dighe:1999bi,Takahashi:2001ep,
Lunardini:2003eh} 
that if the 
propagation of neutrinos is non-adiabatic in the SN matter, then 
they provide us a window to glean into the important 
parameter $\theta_{13}$.  Since both
neutrino and antineutrino fluxes arrive from a SN,
one can in principle also get 
an idea about the neutrino 
mass hierarchy, by observing the difference in the 
neutrino and antineutrino event rates in the detectors. 
The possibility that the neutrinos can in addition undergo earth matter 
effect depending on the location of the SN opens up interesting possibilities 
of studying earth matter effects by comparing the neutrino fluxes 
at different detectors
\cite{Lunardini:2001pb,Takahashi:2001dc,Dighe:2003jg,Dighe:2003vm}. 
This in turn can help   
to distinguish  between the mass hierarchies
\cite{Dighe:1999bi,Lunardini:2001pb,Takahashi:2001dc}. 
In a realistic picture one should consider the 
uncertainties coming from the SN neutrino fluxes 
as discussed in detail in \cite{Lunardini:2003eh}. 
However one can construct variables
through which one can still hope to discover $2-3\sigma$ effects \footnote{
An alternative proposal was discussed in \cite{valle,barger}, 
whereby using the 
known values of the neutrino oscillation parameters 
one could determine the 
SN parameters.}. 

In this paper we do an extensive study of the \kl detector 
towards detection of SN neutrinos.
The dominant reaction for detection of neutrinos 
in \kl is the $\bar\nu_e+p \rightarrow e^{+}+n$.
The liquid scintillation detector material 
also offers the possibility of studying the 
neutrino interactions on the 
Carbon nuclei \cite{uskl}.   
This can proceed via both charged current --
$^{12}{C}(\nu_e,e)^{12}{N}$ and $^{12}{C}({\bar\nu_e,e^{+}})^{12}{B}$ --
and neutral  current -- $^{12}{C}(\nu_x,\nu_x)^{12}{C^*}$. 
The charged current reactions on $^{12}C$ are detected by the delayed
coincidence of the
final state positron/electron with the electron/positron produced by
subsequent $\beta$ decay of the  daughter nuclei.
The event signatures are separated by about tens of millisec in time
and can be tagged. The final state carbon produced
in the neutral current reaction on $^{12}C$
produces a 15.1 MeV $\gamma$ 
-- the detection of which requires large homogeneous detector volume and 
good energy resolution \cite{Cadonati:2000kq}. 
\kl detector satisfies both these criteria.
The low energy threshold of \kl also makes it possible to 
detect the  neutrino-proton elastic scattering 
reaction for the SN neutrinos 
\cite{Beacom:2002hs}.
Neutrinos and antineutrinos of all three flavors participate in this 
neutral current reaction and the event rate is comparable 
and can even be greater than  
the $\bar\nu_e+p$ events.

We explore the information that one can obtain on neutrino oscillation
parameters from the observation of supernova neutrinos through the above
mentioned reactions in KamLAND.
Since the average energies of the $\numu/\nutau(\anumu/\anutau)$
are larger than the average energies of $\nu_e(\bar\nu_e)$,
flavor oscillations cause enrichment of the $\nu_e(\bar\nu_e)$
flux in higher energy neutrinos, thereby
resulting in the {\it hardening} of the corresponding spectra.
Therefore, one expects an enhancement in the CC events
in presence of oscillation. This enhancement could be
particularly  prominent for the CC reactions on $^{12}C$,
for which the energy thresholds are very high
and for the unoscillated $\nu_e$ or $\bar\nu_e$ one
expects very few events.
Since the degree of enhancement depends on the survival probabilities and
hence on the oscillation parameters, especially the mixing angle
$\theta_{13}$ and the neutrino mass hierarchy, the total observed
event rates in \kl can be used to constrain these parameters.
We study the variation of the total event rates
with $\theta_{13}$ and
discuss the limits on $\theta_{13}$ that one can obtain for a
given mass hierarchy.
We also  expound the possibility of differentiating between
the two hierarchies from the observation of energy integrated event rates,
thereby determining the sign of $\Delta m^2_{31}$.
 
The hardening of the  $\nu_e(\bar\nu_e)$ flux
is also manifested in the observed electron/positron
spectrum in the detector. Thus, by analysing the
observed CC spectrum, one may gain information on oscillation
parameters. In addition, if the SN neutrino cross the earth matter
to reach the detector, then the survival probabilities would be
modified due to matter effects inside the earth.
This ``earth regeneration'' effect
is dependent both on the solar neutrino oscillation
parameters, as well as on $\theta_{13}$ and the hierarchy.
Thus the observation of these earth modulations would also give
insight into these parameters.
The earth matter effect is seen as very fast oscillations
superimposed on the slowly varying spectrum of the neutrinos
coming from the SN. Since the variation with energy is fast,
averaging effect due to 
energy resolution of the detector could
wash out the earth effect completely from the energy spectrum.
Therefore one needs a
detector with very good energy resolution. \kl being a
scintillation detector, has a very good energy resolution.
We compare the degree of earth matter effect in the various 
CC reaction channels in \kl. 
The reactions involving the  $^{12}{C}$ are seen to have 
large earth effect induced
fluctuations in the lower energy bins.
However,
with the present volume of 1 kton, \kl lacks the statistics
for drawing precise  constraints from the
$^{12}C$ spectral data.
We present the minimum detector mass required to
detect these modulations at a statistically significant level
in various energy bins. 
We also present the spectrum for a 50 Kton
scintillator detector and discuss the chances of 
probing $\theta_{13}$ and hierarchy.
 
Finally,
we discuss the role of the uncertainties originating from
our lack of knowledge of supernova neutrino fluxes.
The NC reactions (both on Carbon and free proton)
give a measure of the total neutrino
flux coming from SN. 
We introduce the
the ratios of charged and neutral current events
in which
the absolute normalisation uncertainties, e.g.
due to the SN absolute luminosity or the distance of the
supernova from earth,
get canceled. We discuss the merits of these ratios in studying the
oscillation parameters.
We find that these ratios
provide a good handle to probe oscillation parameters particularly when we
consider the neutral current $\nu-p$ scattering reactions.

\section{Neutrino Emission from type-II Supernova }
\label{SN}

A star in an advanced stage of burning has an onion shell structure
with different nuclear reactions going on in each shell.
After millions of years of burning the nuclear reactions stop
inside  the innermost shell (i.e. the core) with matter consisting
mostly of $^{56}$Fe like nuclei.
Once this energy generation in the core stops and if the core has mass
greater than the Chandrasekhar mass, the star undergoes a very fast
gravitational collapse.
During this stage, electron neutrinos are produced through
neutronisation reactions -- electron capture on free protons 
and nuclei. 
During the early stage of the collapse, matter density
of the core remains below the ``neutrino trapping density''
( $\sim$ 10$^{12}$ g/cc),
the neutrinos have mean free path much larger than the radius
of the core and hence escape.
Within a duration of tens of  milliseconds, this collapse compresses
the inner core of the star beyond nuclear matter densities. 
Consequently this inner
core rebounds and creates a shock wave deep inside. This shock wave
propagates radially outward through the star,
gets stalled in the middle due to loss of energy through
nuclear dissociation, eventually gets re-energized
and finally hits the outer
mantle in a few seconds, supplying an explosion energy of a
few times 10$^{51}$ ergs. This is believed to be the cause of the
type II supernova explosion. During the shock phase,
due to high  temperature inside  the core of the star,
thermal $\nu$ and $\bar{\nu}$ are produced in all three flavors
from $e^{+}e^{-}$ pair production and ``urca processes''
\cite{shapiro}.
This production of neutrino pairs stops when neutrino density
becomes high enough so that the inverse process of pair production
balances the direct process. The total energy of the neutrinos from
pair production at the shock phase is much larger than that of the
neutronisation  neutrinos.
Almost all the gravitational
binding energy released due to the collapse is radiated away through these
neutrinos with only a few percent going to the explosion.
These neutrinos inside the core of the
star are in  equilibrium with the ambient matter density and
their energy distributions are close to Fermi-Dirac, with the 
characteristic temperature given by the 
temperature of last scattering.
This is seen through
simulations and through the analysis of
SN1987A neutrinos.
Out of all neutrinos, $\nu_\mu$, $\nu_\tau$
and their antiparticles interact with matter only through neutral
current, whereas $\nu_e$ and $\bar\nu_e$ have
both charged current and neutral current interactions.
Since matter inside the SN is neutron rich, $\nu_e$'s interact more
than $\bar\nu_e$'s. Therefore  $\nu_\mu$, $\nu_\tau$ and their 
antiparticles decouple first and hence have the 
largest temperature, followed by the $\bar\nu_e$, while 
$\nu_e$ decouple last and have the smallest temperature.
In this paper 
we consider the detection of neutrinos arriving at earth
from the cooling phase of the SN burst, and assume that 
all the three types
of neutrino gas  ($\nu_e$, $\bar\nu_e$ and  $\nu_x$,
where $x$ stands for the four species
$\nu_\mu$, $\bar\nu_{\mu}$, $\nu_\tau$
and $\bar\nu_\tau$)
have the Fermi-Dirac energy distributions with  equilibrium  temperatures  
as 11, 16 and 25 MeV respectively 
(see for instance \cite{Langanke:1995he}).
However 
although the 
temperature hierarchy are generally believed to be  
in the above order the exact values of the average energies may differ 
depending on the details of supernova simulations
\cite{Totani:1997vj,Keil:2002in,Raffelt:2003en}. 
Such simulations also indicate that the high energy tail of the 
neutrino spectra from SN is better reproduced by a ``pinched'' 
Fermi-Dirac distribution with a non-zero chemical 
potential \cite{pinch}.
Finally in this paper we have considered that the total binding energy
(3$\times$10$^{53}$ ergs) released in the core collapse of the star
is equally partitioned between various flavors.
But detail simulations may give rise to scenarios where  
the $\nue$ and $\anue$ luminosities can vary between (0.5-2) times 
the $\nu_x$ luminosity \cite{Keil:2002in}. 
The propagation of the neutrinos and their conversion from one flavor
to another depend on matter density profile in the supernova mantle.
A radial profile of 
\bea
\rho(r) = 10^{13} C \left(\frac{10 Km}{r}\right)^3
\eea
with C$\sim$ 1 - 15 provides a good description of the matter
distribution
for $\rho > $  1 g/cc \cite{Lunardini:2001pb,notzold,kp,bbb}.

\section{Calculation of the survival probability}

Neutrinos  and antineutrinos of all flavors 
are produced in a supernova burst. 
Before reaching the detectors, 
the neutrinos travel
through the supernova matter, 
propagate through vacuum and finally may even   
travel through the earth's matter depending on the time of the SN burst.
For the range of neutrino oscillation parameters consistent with  
the solar, \kl reactor and SK atmospheric data, one expects 
large matter enhanced oscillations inside the SN and 
regeneration effects inside the earth.
The neutrino detectors on earth can detect individual 
fluxes of $\nu_e$ and $\bar{\nu}_e$ through the CC reactions  
and the combined fluxes
of all the six neutrino species through the neutral current 
reactions. 
For the case where one has only
active neutrinos, it can be shown that the probabilities involved in 
the calculation of the event rates in the earth bound detectors 
are only the survival probabilities 
$P_{ee}$ and $P_{\bar{e}\bar{e}}$. 

In the three-generation scheme 
\footnote   
{We disregard the positive evidence from LSND experiment \cite{lsnd}.
Study of SN neutrinos in the context of four neutrino species have been done
in \cite{murthy}. We also neglect the CP violating phases which is a 
justified 
approximation as for equal luminosities and temperature of 
the $\mu$ and $\tau$ neutrinos coming from SN, 
the relevant probability at the 
detector is only the electron neutrino survival probability which 
does not depend on the CP phases.}
neutrinos would encounter two resonances in the supernova matter.
The first one corresponds to the higher atmospheric mass
scale at a higher density and 
the second one corresponds to the lower solar mass scale at a lower 
matter density. The higher resonance is controlled by $\Delta m^2_{32}$ 
and $\theta_{13}$ while the lower one is mainly controlled by 
$\Delta m^2_{21}$ and $\theta_{12}$. The transition probability 
depends crucially on the sign of the mass squared difference.
Therefore depending on the sign of 
$\Delta m^2_{32}$, which is still completely unknown,
we can have two possibilities
for the neutrino mass spectrum: Normal Hierarchy (NH) when 
$\Delta m^2_{32} > 0$, or Inverted Hierarchy (IH) when
$\Delta m^2_{32} < 0$. 
We briefly summarize below 
the expressions for the survival probabilities
in the three-generation scheme of the neutrino mass spectrum,
with the current allowed mass and mixing parameters and for both 
normal and inverted hierarchy.
We use the standard Pontecorvo-Maki-Nakagawa-Sakata parametrisation 
for the mixing matrix \cite{upmns}.

\subsection{Normal Mass Hierarchy (NH)}

The  mixing angles for neutrinos in matter for 
normal hierarchy are    
given as \cite{kuo}
\bea 
\tan^22\theta^m_{12}(r) &=&  \label{mix12}
\frac{\Delta m^2_{21}\sin2\theta_{12}}
{\Delta m^2_{21}\cos2\theta_{12} - \cos^2\theta_{13} A(r) } \\
\tan^22\theta^m_{13}(r) &=& \label{mix13}
\frac{\Delta m^2_{31}\sin2\theta_{13}}
{\Delta m^2_{31}\cos2\theta_{13} -  A(r) }
\eea
$A(r)$ being the matter induced term given by
\bea
A(r) &=& 2\sqrt{2}G_FN_An_e(r)E_{\nu}
\eea
where $G_F$ is the Fermi constant, $N_A$ the Avogadro
number, $n_e(r)$ the ambient electron density in the supernova
at radius $r$ and $E_{\nu}$ the energy of the neutrino beam.
Since 
$A >> \Delta m^2_{31} >> \Delta m^2_{21}$ 
at the point of production of the neutrinos,  
from equations
(\ref{mix12}) and (\ref{mix13}) we have, $\theta^m_{12} 
\approx \frac{\pi}{2} \approx \theta^m_{13}$, which means that
neutrinos are created in pure $\nu^m_3$ states and the expression
for the survival probability is
\bea
P_{ee} &=& P_H P_L P^\oplus_{1e} + P_H (1 - P_L)P^\oplus_{2e} 
+ (1 - P_H) P^\oplus_{3e} \label{peenh}
\eea
where
$P_{ie}^\oplus$ are the $\nu_i \rightarrow \nue$ 
transition probabilities and may depend on earth matter effect 
if neutrino cross the earth.
$P_H$ and $P_L$ are the ``jump'' probabilities at the high and low 
density resonance transitions respectively and for these 
we use the standard double exponential forms 
\cite{petcov,lisisn}. 
Although this form derived in \cite{petcov}  was 
in the context of solar neutrinos
and exponential density profile, it was shown in \cite{lisisn}  
that this holds good for generic SN density profiles as well.  

For antineutrinos the expressions (\ref{mix12}) and (\ref{mix13})
hold with the minus sign in the denominators 
replaced by a plus sign.
Then, at the antineutrino production point 
$\bar{\theta}^m_{12} \approx 0 \approx \bar{\theta}^m_{13}$ 
(i.e. they 
are created in the $\bar{\nu}^m_1$ state )
and the 
anti neutrinos the survival probabilities are given by 
\bea
P_{\bar e\bar e} &=& (1 - \bar{P_L}) \bar{P}^\oplus_{1e} 
+  \bar{P_L} \bar{P}^\oplus_{2e} \label{peenhbar}
\eea
where the jump probability $\bar{P_L}$ for the anti neutrinos
is given by the standard expression as given in \cite{lisisn,sandhya}.

\subsection{Inverted Mass Hierarchy (IH)}
\begin{figure}
\centerline{\psfig{figure=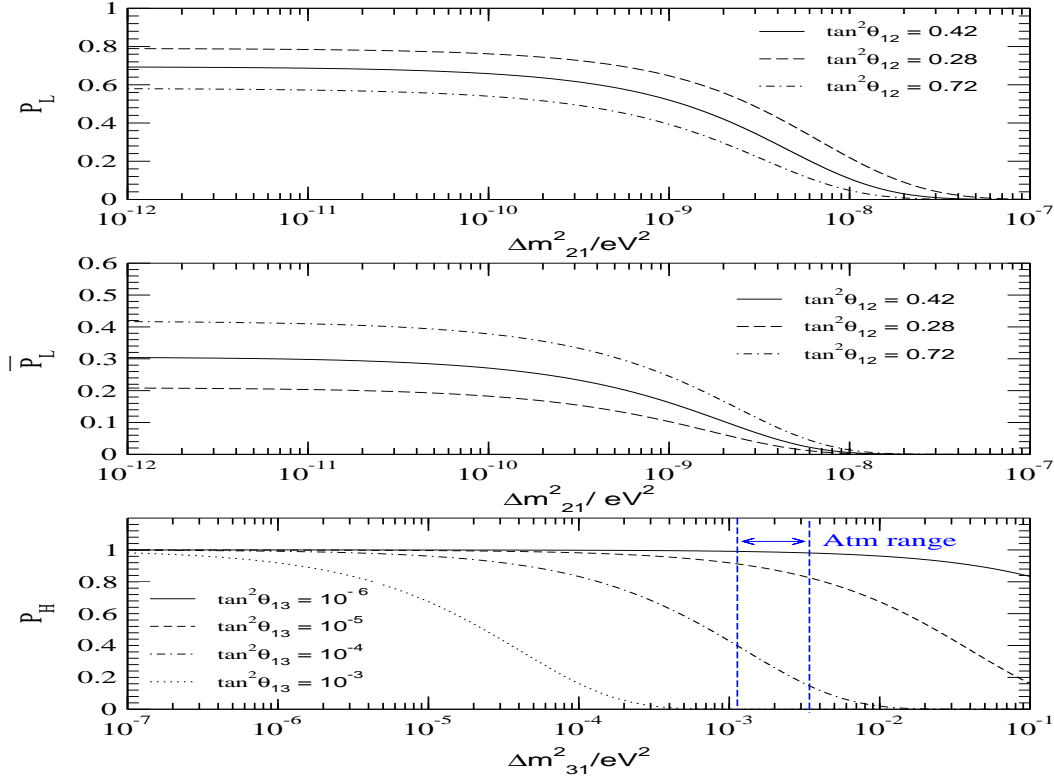,height=12cm,width=15cm}}
\caption{$P_L, \bar{P_L}$ as a function of $\Delta m^2_{21}$ and
$P_H$ as a function of  $\Delta m^2_{31}$ at neutrino energy
$E_\nu$ = 20 MeV}
\label{figprob}
\end{figure}

For inverted mass hierarchy 
($\Delta m^2_{31}  \approx \Delta m^2_{32}  < 0$)
the 12 matter mixing angle 
is same as that for NH 
both for neutrinos and antineutrinos.
The 
13 mixing angle in matter is given by the same expression as for NH with 
the sign of $\Delta m_{31}^2$ reversed.
Then at the production point of the neutrino 
$\theta^m_{12} \approx \frac{\pi}{2}$ and
$\theta^m_{13} \approx 0$ and $\nu_e$ are produced as 
$\nu^m_2$ and the neutrino survival  probability is 
\bea
P_{ee} &=& \label{peeih}
P_L P^\oplus_{1e} + ( 1 - P_L) P^\oplus_{2e}
\eea
whereas at the antineutrino production point
$\bar{\theta}^m_{12} \approx 0$ and
$\bar{\theta}^m_{12} \approx \frac{\pi}{2}$ so that 
$\bar{\nu}_e$ are produced as 
$\nu^m_3$ states and the survival probability is given by
\bea
P_{\bar{e}\bar{e}} &=& \label{peebarih}
(1 - \bar{P_L}) P_H \bar{P}^\oplus_{1e} 
+ \bar{P_L} P_H \bar{P}^\oplus_{2e} 
+ (1 - P_H) \bar{P}^\oplus_{3e}
\eea

\subsection {Dependence of survival  probability on oscillation parameters
and energy}

The dependence of  the jump probabilities 
$P_L$ and $\bar{P_L}$  on the 1-2 mass and mixing
parameters and of  $P_H$ on the 1-3 mass and mixing parameters
has been shown in the figure \ref{figprob}, for 
an illustrative neutrino energy of 20 MeV.
We see from this figure that
for $\Delta m^2_{21} > $ 10$^{-7}$ eV$^2$, $P_L$ = 0 =
$\bar{P_L}$ 
for all values of 
$\theta_{12}$. 
We have checked that this conclusion remains valid 
for all energies of the SN neutrinos.
Since \kl and the solar neutrino data have  
independently confirmed that 
$\Delta m^2_{21}$ lies in the LMA region, we can safely take
$P_L=0$ 
in the expressions for the probabilities.
For the current range of $\Delta m^2_{31}$ 
allowed by the atmospheric neutrino data, 
$P_H$ sharply depends on the value of
$\theta_{13}$. It is therefore useful to consider 
the survival probabilities in three 
limiting range of $\theta_{13}$
\footnote{
In what follows, we keep $\sin^2\theta_{12}$ fixed at 
its current best-fit value of 0.3, and $\Delta m^2_{31}$ fixed 
at $0.002$, the best-fit value presented by the latest re-analysis 
of the SK atmospheric collaboration. The probabilities  
depend on $\Delta m^2_{21}$ if there are earth matter effects, and 
we keep $\Delta m^2_{21}$ fixed at its best-fit value of 
$7.2 \times 10^{-2}$e$V^2$.}
\begin{itemize}
\item {\it The small $\theta_{13}$ range}: For 
$\tan^2\theta_{13} \lsim 10^{-6}$, $P_H =1$ practically over the whole 
relevant energy range of the SN neutrinos, except for very small 
energies. This is the regime of extreme non-adiabatic transitions. 
\item {\it The intermediate $\theta_{13}$ range}:
In the range 10$^{-6} \lsim \tan^2\theta_{13} \lsim 
10^{-3}$, $P_H$ ranges between 
0 and 1, depending on the exact value of $\theta_{13}$ and energy. 
\item {\it The large $\theta_{13}$ range}:
For $\tan^2\theta_{13} \gsim 10^{-3}$ $P_H=0$ for all energies, 
in the entire range of $\Delta m^2_{31}$ 
allowed by the atmospheric neutrino data. 
This is the region of complete 
adiabatic  propagation. 
\end{itemize}

\subsubsection{No earth matter effect} 
\begin{figure}
\centerline{\psfig{figure=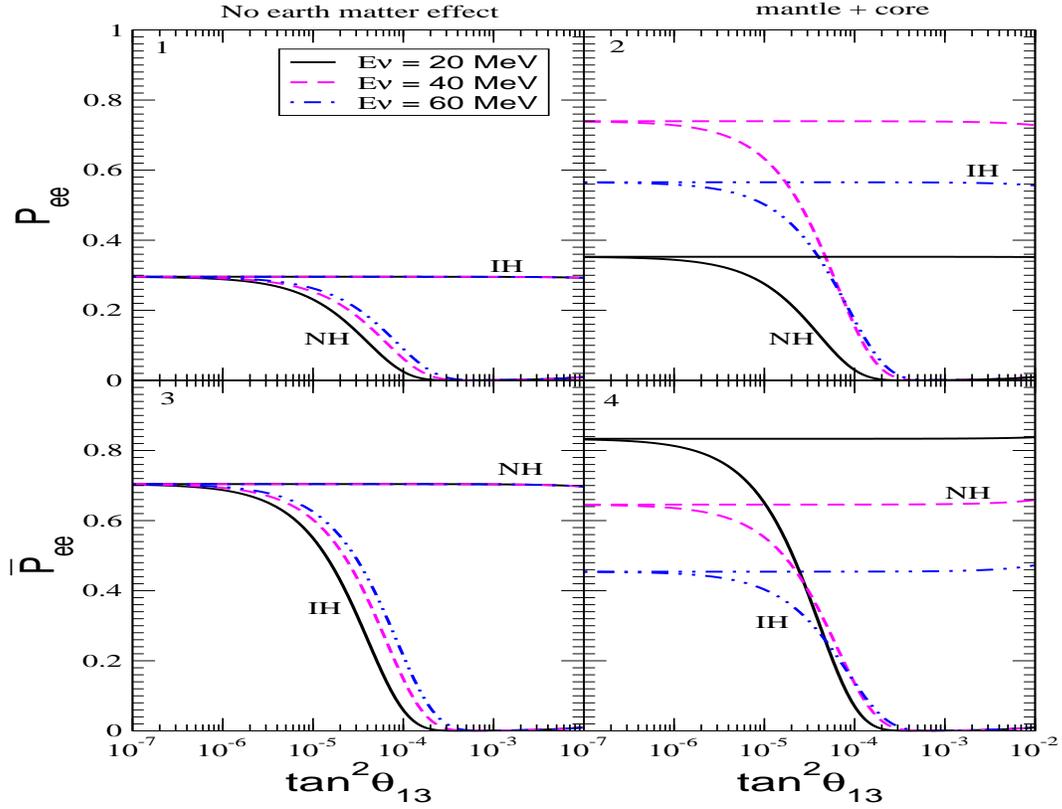,height=12cm,width=15cm}}
\caption{Plot of $P_{ee}^{NH}$, $P_{\bar{e}\bar{e}}^{NH}$,
$P_{ee}^{IH}$ and $P_{\bar{e}\bar{e}}^{IH}$
vs $\tan^2\theta_{13}$ for three different values
of neutrino energy. The left-hand panels show the case for 
neutrinos coming straight from the SN while the right-hand 
panels correspond to the case when they travel earth's mantle and 
core as well.
The other oscillation parameters are kept fixed 
at their current best-fit values.}
\label{figpee13}
\end{figure}

For this case 
\bea
P^\oplus_{1e} &=& \cos^2\theta_{13} \cos^2\theta_{12}\\
P^\oplus_{2e} &=& \cos^2\theta_{13} \sin^2\theta_{12}\\
P^\oplus_{3e} &=& \sin^2\theta_{13}
\eea
Depending on the value of $\theta_{13}$, 
the probabilities assume the following
forms: 
\begin{itemize}
\item
In the range $\tan^2\theta_{13} \lsim 10^{-6}$, $P_H =1$,
$\sin^2\theta_{13} \approx 0$ so that,   
\bea
P_{ee}^{NH}  =  \sin^2\theta_{12}  =  P_{ee}^{IH}\\
P_{\bar{e}\bar{e}}^{NH} = \cos^2\theta_{12} = P_{\bar{e}\bar{e}}^{IH}
\eea
\item
In the range $10^{-6} \lsim \tan^2\theta_{13}\lsim 10^{-3}$, 
$P_H$ is non-zero but $\sin^2\theta_{13}$ can be taken as zero and, 
\bea
P_{ee}^{NH} = P_H \sin^2\theta_{12}
 ;P_{ee}^{IH} =  \sin^2\theta_{12} \\
P_{\bar{e}\bar{e}}^{NH}  =   \cos^2\theta_{12} ; 
P_{\bar{e}\bar{e}}^{IH}  =  P_H \cos^2\theta_{12}
\eea
\item
For $\tan^2\theta_{13} \gsim 10^{-3}$, $P_H=0$, $\sin^2\theta_{13}$ 
can be non-zero  
\footnote{However if we  restrict ourselves up to the CHOOZ bound
then $\sin^2\theta_{13}$ stays very small.}
and the probabilities are, 
\bea
P_{ee}^{NH} = \sin^2\theta_{13} ~~~;~~~
P_{ee}^{IH}  =  \cos^2\theta_{13}\sin^2\theta_{12}  
\\
P_{\bar{e}\bar{e}}^{NH}  =   \cos^2\theta_{13}\cos^2\theta_{12}~~~;~~~
P_{\bar{e}\bar{e}}^{IH}  =  \sin^2\theta_{13}
\eea 
\end{itemize}
The dependence of the probabilities on $\theta_{13}$ 
for the case of no earth matter effects 
are shown in the left-hand panels of  
figure \ref{figpee13}, for three different values of 
the neutrino energy.  
From the figure \ref{figpee13} and the 
equations for the probabilities at various limits 
we note the following points: 
\begin{enumerate}
\item
In the small $\theta_{13}$ regime ($\tan^2\theta_{13} \lsim 10^{-6}$) 
the survival probabilities for both $\nue$ and $\anue$ are
independent of the 
value of $\theta_{13}$ and neutrino energy. 
The flavor conversion is governed only by the solar mixing angle 
$\theta_{12}$, and 
one cannot distinguish between normal and inverted hierarchies 
in either the neutrino or the antineutrino channel.

\item
For IH(NH) the survival probabilities for $\nue(\anue)$
do not depend on $P_H$. Hence they
remain largely independent of the 
value of $\theta_{13}$ and the neutrino energy 
and depend mainly 
on the solar mixing angle $\theta_{12}$ in the entire 
oscillation parameter range 
considered in this paper.
Dependence on $\theta_{13}$ comes  
only for NH(IH) in the $\nue(\anue)$ channel.

\item
In the intermediate $\theta_{13}$ range
($10^{-6} \lsim  \tan^2\theta_{13} \lsim 10^{-3}$)
and with NH(IH) as the true mass hierarchy,
the neutrino(anti-neutrino)
survival probabilities  
depend sharply on the value of $\theta_{13}$ and neutrino energy, 
through the dependence of $P_H$ on these variables.

\item In the large $\theta_{13}$ range ($ \tan^2\theta_{13} \gsim 10^{-3}$ )
and with NH(IH) as the true mass hierarchy,
the neutrino(anti-neutrino)
survival probabilities
are very small, 
within the CHOOZ limit of $\sin^2\theta_{13}$, and  
do not depend on energy. 

\item In the case of $\nue(\anue)$ for NH(IH), the survival 
probabilities decrease from $\sin^2\theta_{12}(\cos^2\theta_{12})$ 
to $\sim \sin^2\theta_{13}(\sin^2\theta_{13})$, 
as $\tan^2\theta_{13}$ changes from $\sim 10^{-6}$ to 
$\sim 10^{-3}$. Since the best-fit value of the solar mixing 
angle is $\sin^2\theta_{12} \approx 0.3$, \peebarih has a much 
sharper decrease with  $\tan^2\theta_{13}$ than \peenh. Therefore the 
$\anue$ events could be expected to have better sensitivity to 
$\theta_{13}$ on this count.

\item 
In the range
$10^{-6} \lsim \tan^2\theta_{13}\lsim 10^{-3}$, $P_{ee}^{NH}$
and $P_{\bar{e}\bar{e}}^{IH}$
depend on $\Delta m^2_{31}$
through $P_H$  but 
over the currently allowed
atmospheric range of this parameter, 
the dependence is not very strong.

\item 
The survival probabilities in general depend on the value of 
$\theta_{12}$, excepting \peenh and \peebarih in the large $\theta_{13}$ 
regime, but there is no dependence of the probabilities on
$\Delta m^2_{21}$.

\end{enumerate}

\subsubsection{Including the earth matter effect}
For this case 
\bea
P^\oplus_{1e} &=& \cos^2\theta_{13}|A^\oplus_{1e}(2gen)|^2 
\\
P^\oplus_{2e} &=& \cos^2\theta_{13}|A^\oplus_{2e}(2gen)|^2 
\\
P^\oplus_{3e} &=& \sin^2\theta_{13}
\eea
$A^\oplus_{1e}$(2gen) and $A^\oplus_{2e}$(2gen)
are the $\nu_i \rightarrow \nue$ transition amplitudes 
inside earth for
two generations \cite{chitre,Petcov:1998su},  with
matter term $A(r)$ replaced by $A(r)\cos^2\theta_{13}$.
Since the supernova neutrinos have typical energy of 
$\sim$ 10 MeV the matter potential in the earth 
is $<<$ the high mass scale $\Delta m^2_{31}$.  However, 
oscillations due to the $\Delta m^2_{21}$ 
scale are affected by the earth matter effect. 

\begin{itemize}
\item
For $\tan^2\theta_{13} \lsim 10^{-6}$, $P_H = 1$, $\sin^2\theta_{13}$ 
can be set to zero and the probabilities are, 
\bea
P_{ee}^{NH} = P^\oplus_{2e}(2gen)
= P_{ee}^{IH}\label{small13pe}\\
P_{\bar{e}\bar{e}}^{NH} =  \bar{P}^\oplus_{1e}(2gen)
= P_{\bar{e}\bar{e}}^{IH} 
\label{small13pae}
\eea

\item
For $10^{-6} \lsim \tan^2\theta_{13} \lsim 10^-3$ the probabilities 
can be written as, 
\bea
P_{ee}^{NH}  \approx  P_H \ptoe ~~~;~~~
P_{ee}^{IH}  \approx  \ptoe \\
P_{\bar{e}\bar{e}}^{NH}  \approx  \poneeb ~~~;~~~
P_{\bar{e}\bar{e}}^{IH}  \approx  P_H \ponee
\eea
where we have used $\sin^2\theta_{13}=0$ which is a good
approximation in this range. 

\item
For $\tan^2\theta_{13} \gsim 10^{-3}$, $\sin^2\theta_{13}$ may be
small but non-zero in the range allowed by the CHOOZ results.
However as we have seen before that $P_H =0$ and the probabilities are,
\bea
P_{ee}^{NH}  =  \sin^2\theta_{13} ~~~;~~~
P_{ee}^{IH}  =  \ptoe \\
P_{\bar{e}\bar{e}}^{NH}  =  \p1eboplus ~~~;~~~
P_{\bar{e}\bar{e}}^{IH}  =  \sin^2\theta_{13}
\eea
\end{itemize}
We can define the following  quantities valid for all values of 
$\theta_{13}$: 
\bea
R_{ee}^{NH}  =  P_H \cos^2\theta_{13} f_{reg} ~~~;~~~
R_{ee}^{IH}   =  \cacsq f_{reg} \label{ree}\\
R_{\bar{e}\bar{e}}^{NH}  =  - \cacsq \bar{f}_{reg} ~~~;~~~
R_{\bar{e}\bar{e}}^{IH}  =  - P_H \cacsq \bar{f}_{reg}
\label{raeae}
\eea 
where $R_{ii}^{x} = P_{ii}^x$(with earth effect) - 
$P_{ii}^x$(without earth effect), 
i can be $e$ or $\bar{e}$ and $x$ can be NH or IH and 
\bea
f_{reg} & = & |A_{2e}^{\oplus}|^2 - \sabsq 
\label{freg}
\\
\bar{f}_{reg} & = & |\bar{A}_{2e}^{\oplus}|^2 - \sabsq
\label{frega}
\eea
The 
$f_{reg}$ and $\bar{f}_{reg}$ 
carry the information about the regeneration effects 
inside the earth and are 
dependent on the neutrino energy,
$\theta_{12}$ and $\Delta m^2_{21}$.

From Eqs. (\ref{ree}) and (\ref{raeae}) we see that while
$ R_{ee}^{IH} $ and $R_{\bar{e}\bar{e}}^{NH}$ are independent 
of $P_H$, 
$R_{ee}^{NH} $ and $  R_{\bar{e}\bar{e}}^{IH}$ are linearly 
proportional to it. Thus, while the $\anue(\nue)$ spectrum
will have an energy (and $\theta_{12}$ and $\Delta m^2_{21}$) 
dependent, but largely $\theta_{13}$ independent modulation induced by 
$\bar{f}_{reg}(f_{reg})$ for NH(IH), for IH(NH) the 
degree of earth matter 
effects will depend sharply 
on the value of $\theta_{13}$. In particular, 
when $P_H=1$ the earth effects will be same for both the hierarchies.
However, as $\theta_{13}$ increases, $P_H$ decreases, resulting in 
smaller earth regeneration effect in the $\anue(\nue)$ spectrum 
for IH(NH). The earth effect completely vanishes in the 
$\anue(\nue)$ channel for IH(NH) 
when $P_H=0$ for $\tan^2\theta_{13}\gsim 10^{-3}$. 

We show the impact of earth matter effect on the survival probabilities 
in the right-hand panels of figure \ref{figpee13}.
From the expressions above and figure  \ref{figpee13}
we note the following points in this case: 
\begin{enumerate} 
\item
In the small $\theta_{13}$ regime ($\tan^2\theta_{13} \lsim 10^{-6}$), 
the probabilities are 
again independent of $\theta_{13}$ but they are not independent of energy
anymore, 
as the earth matter effect introduces some energy dependence. 
Apart from $\theta_{12}$ the probabilities in this regime are 
now also functions of $\Delta m^2_{21}$ through the dependence 
on $f_{reg}$ and $\bar{f}_{reg}$.
However  as is clear from figure \ref{figpee13} and
Eqs. (\ref{small13pe}) and (\ref{small13pae}) 
the  probabilities are same for both 
hierarchies and introduction of earth effect does not help in lifting the 
degeneracy of the probabilities between normal and inverted hierarchies. 

\item
As can be seen clearly from figure \ref{figpee13},
\peebarnh and \peeih 
become energy dependent 
with the inclusion of earth matter effect 
but for a fixed energy
there is no dependence of these probabilities on $\theta_{13}$. 

\item 
In the intermediate $\theta_{13}$ range
($10^{-6} \lsim  \tan^2\theta_{13} \lsim 10^{-3}$)
for the $\anue(\nue)$ channel
with IH(NH) as the true mass hierarchy,
the $\theta_{13}$ dependence is governed by $P_H$.  
Whereas the energy dependence of the 
the neutrino(anti-neutrino)
survival probabilities 
is due to  both  $P_H$ 
and the earth regeneration factor $\bar f_{reg}(f_{reg})$. 
Since the amount of  earth regeneration 
for these channels 
depends on $P_H$ (cf. Eq. (\ref{raeae}))
the degree of energy dependence 
is different for different ranges of $\theta_{13}$ -- there being more 
earth matter induced energy variation at smaller $\theta_{13}$ as
can be seen in figure \ref{figpee13}. 

\item For large values of $\theta_{13}$ ($ \tan^2\theta_{13} \gsim 10^{-3}$ )
\peenh and \peebarih 
are given by the same expressions as the 
no earth effect case and do not depend on energy
or earth matter effects. But \peeih and \peebarnh 
depend on earth matter effect and energy.  


\end{enumerate}

\section{Supernova neutrino signal in the \kl detector}

\subsection{Energy integrated event rates in \kl}
Neutrinos from a SN can be observed in \kl
through their charged and neutral  current reactions with the
detector scintillator material. The total number of charged current
events triggered by the SN $\anue$ flux arriving at \kl
can be expressed as,
\bea
R^{CC} &=&  \nonumber
\eta N \times \int dE_{vis} \int dE_\nu \sigma(E_\nu) R(E_{vis},E_\nu)
\sigma_{CC} (E_{\nu})
\nonumber \\
& & \times \Big{[}F_{\bar e}(E_{\nu}) P_{\bar e\bar e}
(E_{\nu}) + F_{\bar x}(E_{\nu})
(1 - P_{\bar e \bar e}(E_{\nu}) ) \Big{]}
\label{rcc}
\eea
where
N is the number of target nuclei/nucleons, $\eta$ is the
detection efficiency, $\sigma_{CC} (E_{\nu})$
are the corresponding CC cross-sections for the
$\anue$ and
$E_{vis}$ is the measured {\it visible} energy of the emitted
positron, when the true visible energy, $E_{vis}^T$ is related to
the antineutrino energy $E_{\nu}$ as,
$ E_{vis}^T \cong E_{\nu}-0.78$ MeV.
$R(E_{vis},E_{e^+})$ is the Gaussian energy resolution function of the
detector given by,
\bea
R(E_{vis},E_{e^+}) &=&
\frac{1}{\sqrt{2\pi\sigma_0^2}}\exp\left( - \frac{(E_{vis} - E_{e^+}
-m_e)^2}
{2\sigma_0^2}   \right)
\eea
with
$\sigma_0(E)$/E = 7.5$\%/\sqrt(E)$ as given by the KamLAND
collaboration \cite{kl}.
$F_{\bar e}$ is the unoscillated electron antineutrino
flux produced in the SN and $F_{\bar x}$ stands for the flux of any of
the other antineutrino flavors, $\anumu$ or $\anutau$.
The expression for the
neutrino flux of species $\alpha$, arriving at the detector is
given by
\bea
F_{\alpha} (E_\nu, L_{\alpha}, T_{\alpha}, d )
&=& \frac{L_{\alpha}}{4 \pi d^2} \times
\frac{E_{\nu}^2} {5.68 T_{\alpha}^4
\left[ \exp(\frac{E_{\nu}}{T_{\alpha}} )+ 1 \right] }
\label{fl}
\eea
where $L_{\alpha}$ is the total SN energy released in
the neutrino species $\nu_{\alpha}$, $T_{\alpha}$ is the
 temperature of the $\nu_{\alpha}$ gas at the neutrino
sphere and $d$ is the distance of the supernova from earth.
In Eq.(\ref{fl}) we assume a pure Fermi-Dirac spectrum for the
neutrinos. As discussed before in section \ref{SN}, the actual SN neutrino
spectrum may deviate from a perfect black body. This
deviation can be accounted for by introducing in the neutrino
distribution function a ``pinching factor'' $\eta$, which is 
a dimensionless parameter defined as $\mu_\alpha/T_\alpha$, where 
$\mu_\alpha$ and $T_\alpha$ are the 
chemical potential and the temperature of the neutrino 
species $\nu_\alpha$ \cite{pinch}. The pinching factor 
has the effect of suppressing the high energy end of the
neutrino spectrum \footnote{Consequences of a 
a small deviation from Fermi-Dirac in the high
energy tail, for neutrino oscillation parameters,
have been discussed for instance in \cite{Lunardini:2003eh}.}.

The number of CC events induced by the SN $\nue$ in \kl
is given by the same expression Eq.(\ref{rcc}), but with the antineutrino
fluxes $F_{\bar e}$ and $F_{\bar x}$ replaced by the neutrino fluxes
$F_{{e}}$ and $F_{{x}}$,  $P_{\bar e\bar e}$ replaced by 
$P_{ee}$ and with the corresponding CC interaction cross-section 
for the neutrinos.
The expression for the neutral current events in the detector is
given by
\bea
R^{NC} &=&   \nonumber
\eta N\int dE_{\nu} \\
& & \times \left[ \sigma_{NC} (E_\nu)
\Big{(} F_e(E_{\nu}) + 2 F_x(E_{\nu}) \Big{)} + \bar{\sigma}_{NC} (E_\nu)
\Big{(}F_{\bar {e}}(E_{\nu}) + 2 F_{\bar {x}}(E_{\nu})\Big{)} \right]
\label{rnc}
\eea
where $\sigma_{NC}(E_\nu)$ and $\bar{\sigma}_{NC} (E_\nu) $ are the
NC cross-sections corresponding to neutrinos and antineutrinos
respectively and all other variable are as defined before.
\subsection{Detection reactions in \kl}

\kl is an isoparrafin based  liquid scintillator
(80\% paraffin oil and 20\% pseudodocumene) detector
with a typical composition of $C_n H_{2n}$.
The main CC detection reactions in \kl are:
\begin{enumerate}

\item The CC capture of $\bar\nu_e$ on free protons:
\bea
\bar{\nu_e}+p \rightarrow e^+ + n
\eea
The threshold $\anue$ energy for this reaction
is 1.8 MeV. The released positron annihilates with an
ambient electron to produce two gamma rays.
The final state neutron
thermalises in about 180 $\mu sec$ to produce a 2.2 MeV $\gamma$ ray.
The delayed coincidence of
these $\gamma$s gives an event grossly free from backgrounds.
\footnote{This is the reaction used by \kl to detect the
reactor antineutrinos}.

\item
The CC capture of $\bar\nu_e$ on $ ^{12}C$:
\bea
\bar{\nu_e} + ^{12}C \rightarrow  ^{12}B + e^+\\
^{12}B \rightarrow ^{12}C  + e^- + \bar{\nu_e}
\eea
The threshold antineutrino energy for the capture on
$ ^{12}C$ is 14.39 MeV. The $\beta$ unstable $^{12}B$
eventually decays with a half life of $\tau_{1/2}=20.20$ ms.
This signal is also recorded by the delayed coincidence
of the $\gamma$s produced from annihilation of
e$^+$ produced in the final state of the first reaction,
followed by the annihilation of e$^-$
produced in the final state of the second reaction.

\item
The CC capture of $\nu_e$ on $ ^{12}C$:
\bea
\nu_e + ^{12}C \rightarrow  ^{12}N + e^-\\
^{12}N  \rightarrow  ^{12}C + e^+ + \nu_e
\eea
The threshold neutrino energy for capture on
$ ^{12}C$ is 17.34 MeV and the $\beta$ unstable
$^{12}N $ created in this process decays with a
half life of $\tau_{1/2}= 11.00$ ms. The signal is
again recorded by the
delayed coincidence
of the two $\gamma$s produced from annihilation of the
e$^+$ and e$^-$ respectively, produced in the final states
of the two above reactions.
\end{enumerate}

The main NC reactions by which \kl can observe SN neutrinos are:
\begin{enumerate}
\item
The neutrino-proton elastic scattering:
\beq
\nu + p \rightarrow \nu + p
\eeq
The scattered protons have kinetic energies of $\sim$ MeV and hence
cannot be observed in the high-threshold Cerenkov detectors.
However in a scintillator detector like KamLAND, the
energy deposition due to the protons can be
observed in spite of the quenching of the proton energy due to ionisation
\cite{Beacom:2002hs}. The cross-section for $\nue$ scattering is same as that
of scattering by $\nu_x$, while the
cross-section for $\anue$ scattering is same as that
of scattering by $\nu_{\bar x}$ \cite{Beacom:2002hs}. But the scattering
cross-section of the particles is different from the scattering
cross-section of the anti-particles. However under the approximation
of $E_\nu/M_p <<1$, where $M_p$ is the proton mass, even the
particle and anti-particle cross-sections can be taken as
almost equal.
\item
Neutral Current break-up of $ ^{12}C$:
\bea
\nu(\bar{\nu}) + ^{12}C \rightarrow  ^{12}C^*
+ \nu^\prime(\bar{\nu^\prime})  \\
^{12}C^* \rightarrow  ^{12}C + \gamma (15.11 MeV)
\eea
The threshold energy for this reaction is
15.11 MeV. The detection of the 15.11 MeV $\gamma$ released by the
de-excitation of the $^{12}C^*$ gives an unambiguous signal of
the supernova neutrinos, virtually free from all backgrounds.
\end{enumerate}

All the above reactions have distinctive signatures, allowing one to count
the event rates separately. The CC reactions can also measure the
energy spectrum of the final state electrons/positrons.
In the $\nu-p$ NC scattering reaction, the scattered proton
carries the energy information and hence this reaction also
has sensitivity to the spectrum of the incoming neutrinos.
However for the NC break-up of $ ^{12}C$,
the emitted photon has no information
about the neutrino energy and this is the only reaction in
\kl with no energy sensitivity. We take the $\anue+p$ absorption
cross-section from \cite{vogel_anup}, the $\nu-p$ elastic scattering
cross-section from \cite{Beacom:2002hs} and the $^{12}C$ reaction
cross-sections from \cite{vogel-12c}.

Finally a note on the expected backgrounds in KamLAND.
The Kamioka mine has a rock overburden of 2700 m.w.e producing
0.34 Hz of cosmic-ray muons.
Since the SN neutrino signal lasts for about 10 s,
the signal for
the $\bar{\nu_e}+p$ events is $\sim$ 300/10 = 30 Hz, while that for the
Carbon events are 10/10 = 1 Hz.
Thus the signal is greater than the cosmic ray backgrounds.
Since the CC reactions are observed through delayed coincidence
techniques, therefore they are grossly free of background problems.
The only problem may come from low-energy radioactive backgrounds
for the $\nu-p$ scattering reaction. However Borexino has already demonstrated
that it is possible to understand this background. If 
the KamLAND collaboration want to
pursue the low-energy solar neutrino program then they also have to
understand this background properly.
However it should be noted that since the SN neutrino signals
last for about 10 s and hence is a time varying signal
the background for these is almost
3 times lower than the solar neutrino background \cite{Beacom:2002hs}.

\subsection{Supernova neutrino signal in \kl}
\begin{figure}
\centerline{\psfig{figure=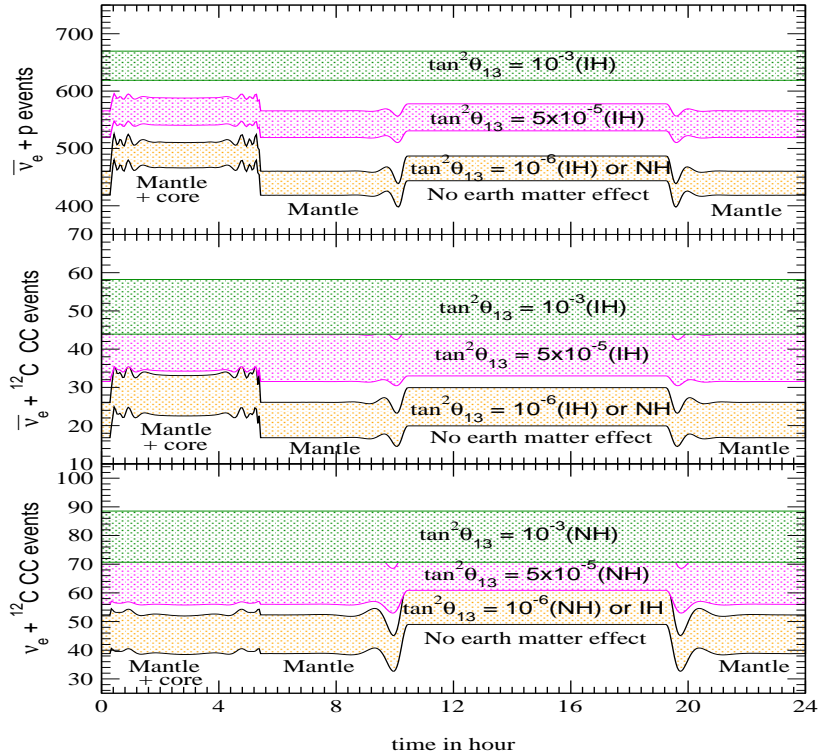,height=12cm,width=12cm}}
\caption{Plot of the event rates with 1$\sigma$ statistical error
bands in  the different CC reaction in
\kl as a function of
time of  occurrence of the Supernova
for different values of $\theta_{13}$. The curves are drawn with
$\Delta m^2_{21}$ and $\theta_{12}$ fixed at the best-fit value from the
global solar and \kl data. See text for details. }
\label{fignuebarpsk}
\end{figure}

\begin{table}
\begin{tabular}{||c||c|c|c|c||c|c|c|c||c|c|c|c||c||c||}
\hline
\hline
& \multicolumn{4}{|c||}{$\bar{\nu}_e$+p$\rightarrow$n+e$^+$}
& \multicolumn{4}{|c||}{$\nu_e$+$^{12}$C$\rightarrow^{12}$N+e$^+$}
& \multicolumn{4}{|c||}{$\bar{\nu}_e$+$^{12}$C$\rightarrow^{12}$B+e$^+$}
& & \\
\cline{2-13}
\cline{2-13}
 & \multicolumn{2}{|c|} {no}
 & \multicolumn{2}{|c||}{with}
 & \multicolumn{2}{|c|}{no}
 & \multicolumn{2}{|c||}{with}
 & \multicolumn{2}{|c|}{no}
 & \multicolumn{2}{|c||}{with}
 & $\nu$+$^{12}$C & $\nu$+p \\
$\tan^2\theta_{13}$ &\multicolumn{2}{|c|} {earth}
     &\multicolumn{2}{|c||} {earth}
     &\multicolumn{2}{|c|} {earth}
     &\multicolumn{2}{|c||} {earth}
     &\multicolumn{2}{|c|} {earth}
     &\multicolumn{2}{|c||} {earth}
& $\rightarrow^{12}$C$^*$+$\nu$
& scatt.\\
 & \multicolumn{2}{|c|} {effect}
 & \multicolumn{2}{|c||} {effect}
 & \multicolumn{2}{|c|} {effect}
 & \multicolumn{2}{|c||} {effect}
 & \multicolumn{2}{|c|} {effect}
 & \multicolumn{2}{|c||} {effect}
& & \\
\cline{2-13}
 & NH & IH & NH & IH & NH & IH & NH & IH & NH & IH & NH & IH & & \\
\hline
10$^{-6}$          & 463 & 463 & 487 & 487 & 56 & 56 & 45 & 45 & 25 & 25 & 27 &
27 && \\
\cline{1-13}
5$\times$10$^{-5}$ & 463& 554 & 487 & 564 & 69 & 56 & 64 & 45 & 25 & 39 & 27 & 
41 & 44 & 676 \\
\cline{1-13}
10$^{-3}$          & 463 & 644 & 487 & 644 & 80 & 56 & 80 & 45 & 25 & 51 & 27 & 51 && \\
\cline{1-13}
No osc. & \multicolumn{4}{|c||} {383} &  \multicolumn{4}{|c||}{2.6}
 & \multicolumn{4}{|c||}{13} && \\
\hline
\end{tabular}
\caption{\label{table}
The number of events for the different reactions in \kl
in presence and absence of oscillation.
For the case with oscillation we present the events for three different values
of $\theta_{13}$ and for both normal and inverted hierarchy.
The NC events are unaffected by oscillation.}
\end{table}

The total number of CC and NC event rates in KamLAND
for the different reactions
listed in the previous subsection, can be obtained
from the Eqs. (\ref{rcc}) and (\ref{rnc}).
In Table \ref{table} we present the total number of events
expected in 1 kton of detector mass with 100\% assumed efficiency,
induced by a galactic supernova at a distance of 10 kpc
and located at the galactic centre
with $\delta_s= -28.9^\circ$ (see Appendix). 
The lowest row
in Table \ref{table}
gives the number of events for the case of no oscillations.
Also given are the changed number of events in presence of
oscillation, for both normal and inverted mass hierarchy,
and for three different  representative 
values of $\theta_{13}$, chosen from the three
interesting ranges
discussed in the previous section. For events with oscillations,
we present results obtained both with and without earth matter effects.
The earth matter effects depend on the trajectory of the SN neutrino
beam through the earth and hence on
the direction of the SN from the detector.
The direction of the SN with respect to the detector depends
on $\delta_s$,  
as well as the time during the day
when the supernova explodes.
In Table \ref{table} the
number of events presented with earth matter effect 
correspond  to a typical
time instant for which the neutrinos cross both
the earth's mantle and core.
Figure \ref{fignuebarpsk} shows the variation of the SN signal
in the detector as a function of the time at which the supernova
neutrinos arrive. The number of events for the
CC reactions in \kl are shown in figure \ref{fignuebarpsk}
for the three typical values of $\theta_{13}$.
For $\tan^2\theta_{13} = 10^{-6}$ the $\anue$ and $\nue$ events 
correspond to both Normal and Inverted hierarchy.
As we increase $\theta_{13}$ 
the number of $\anue$($\nue$) events for NH(IH) does not change
and continue to be given by the same band. 
The events for higher $\theta_{13}$ 
values correspond to the case of inverted hierarchy for the
case of $\anue$ events and normal hierarchy for the case of the
$\nue$ CC capture on $^{12}C$.
We refer the reader to Appendix A for
the details of the geometry of the neutrino trajectory inside the earth,
which is used for the calculation of the number of events in the
presence of earth matter effects throughout this paper.

The no oscillation number of events
for the $\bar\nu_e+p$ absorption is 383 events while the
$\nu-p$ elastic scattering gives 676 events.
The cross-section for the $\nu-p$ NC
scattering reaction is about 4 times lower
than the $\bar\nu_e + p$ CC absorption reaction. But
since all the six neutrino species take part in the former
and since the $\nu_\mu$ and $\nu_\tau$ are more energetic than $\bar\nu_e$,
one gets a larger number of $\nu-p$
elastic scattering events for the case of no oscillation.
For the $\nu-p$ events presented in Table \ref{table}
we have ignored energy threshold effects.
If we assume a threshold of 0.28 MeV, determined mainly by the
radioactive backgrounds, then the number of no oscillation
events for the $\nu-p$ elastic scattering reduces to 270.
The energy threshold for the
$^{12}{C}$ reactions are very high
and so there are very few events in absence of oscillations.

Neutrino flavor oscillations convert higher energy
$\numu/\nutau$($\anumu/\anutau$) into $\nue(\anue$), resulting in
an enhancement of the charged current events.
The $\anue+p$ CC events increase by a factor of about 1.2 -- 1.7,
depending on the value of $\theta_{13}$ and the neutrino mass
hierarchy.
The number of $^{12}{C}$ CC events increase
dramatically with the introduction of neutrino flavor conversion.
The $\nue-^{12}{C}$ events increase by a factor of $\sim 17 - 30$,
while $\anue-^{12}{C}$ events could see an increase of
$\sim 2-4$ times, over the no oscillation expected rate.
From the Table \ref{table} we see that
the extent of the increase in the number of events
for a particular reaction channel depends on
the value of $\theta_{13}$ and the neutrino mass hierarchy.
However, for a given $\theta_{13}$ and hierarchy,
the difference between
the degree of enhancement as a result of oscillations
for different reactions can be
understood in terms of:
(i) relative change in the average energy of the $\nue/\anue$
flux due to flavor oscillations and
(ii) degree of dependence of the CC cross-section on
the neutrino energy. The relative change in the average energy
due to oscillations for $\nue$ 
is much larger than that for $\anue$. This explains
why the $\nue$ events show a larger enhancement than
the $\anue$ events. Between the $\anue$ events, we
note that the relative increase for $\anue-^{12}{C}$ reaction
is more than for $\anue+p$. This is because the cross-section
for the $\anue-^{12}{C}$ has a sharper dependence on energy than
the $\anue+p$ cross-section.
The neutral current event rate is invariant under
neutrino oscillations since the NC reactions are flavor blind.


\section{Probing $\theta_{13}$ and mass hierarchy}

\subsection{Using total event rates}

From the Table \ref{table} 
and figure \ref{fignuebarpsk}  we observe that the increase in number of the 
$\nu_e$ events for normal hierarchy
and the $\bar\nu_e$ events for inverted hierarchy 
depend on the value of $\theta_{13}$.
Therefore the total number of events observed may be used 
to probe $\theta_{13}$ and hierarchy. 

\begin{figure}
\centerline{\psfig{figure=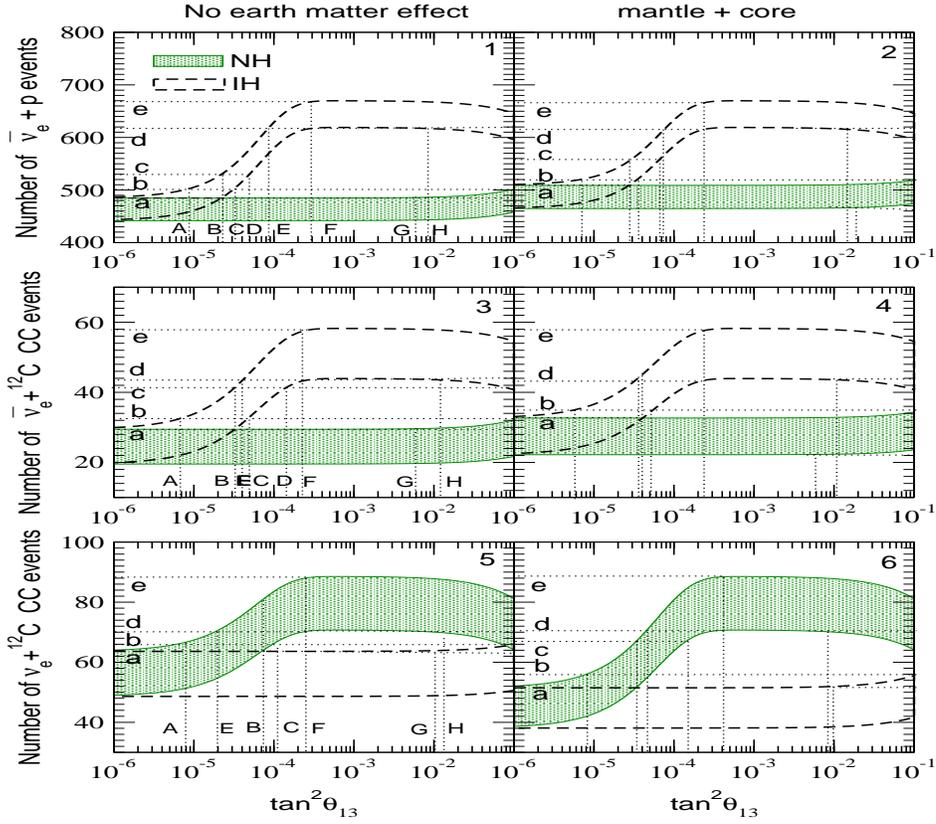,height=12cm,width=14cm}}
\caption{Plot of the CC event rates as a function of
$\theta_{13}$ for NH and IH for KamLAND. The 1$\sigma$ statistical error 
band is also presented. See text for details.}
\label{figrate_stat}
\end{figure}



\begin{table}
\begin{center}
\begin{tabular}{|c|c|c|}
\hline
Reactions & Range of observed events &  conclusions drawn\\
\hline
\hline
$\bar{\nu}_e$+p$\rightarrow$n+e$^+$
 & N$<$ 485(a) & NH with no bound on $\theta_{13}$  \\
 &                   & OR IH with $\t<2.3\times$10$^{-5}$(B) \\
\cline{2-3}
 & 485(a)$<$N$<$501(b) & NH with $\t>6.2\times10^{-3}$(G)\\
 &     & OR IH with $\t<3.4\times10^{-5}$(C) \\
\cline{2-3}
 & 501(b)$<$N$<$530(c) &
 IH with $8.7\times10^{-6}(A)<\t<4.7\times10^{-5}(D)$\\
\cline{2-3}
 & 530(c)$<$N$<$615(d) &
IH with $2.3\times10^{-5}(B)<\t<2.9\times10^{-4}(F)$\\
 & & OR IH with $\t>8.4\times10^{-3}(H)$\\
\cline{2-3}
 & N$>$615(d) & IH with $\t>8.6\times10^{-5}$(E) \\
\hline
\hline
$\bar{\nu}_e$+$^{12}$C$\rightarrow^{12}$B+e$^+$
 & N$<$ 29(a) & NH with no bound on $\theta_{13}$  \\
 &                   & OR IH with $\t<3.3\times$10$^{-5}$(B) \\
\cline{2-3}
 & 29(a)$<$N$<$33(b) & NH with $\t>5.6\times10^{-3}$(G)\\
 &     & OR IH with $\t<4.9\times10^{-5}$(C) \\
\cline{2-3}
 & 33(b)$<$N$<$41(c) &
 IH with $6.8\times10^{-6}(A)<\t<1.4\times10^{-4}(D)$\\
\cline{2-3}
 & 41(c)$<$N$<$44(d) &
IH with $3.3\times10^{-5}(B)<\t<2.3\times10^{-4}(F)$\\
 & & OR IH with $\t>1.2\times10^{-2}(H)$\\
\cline{2-3}
 & N$>$44(d) & IH with $\t>4.2\times10^{-5}$(E) \\
\hline
\hline
$\nu_e$+$^{12}$C$\rightarrow^{12}$N+e$^+$
 & N$<$ 63(a) & IH with no bound on $\theta_{13}$ \\
 &                   & OR NH with $\t<7.4\times$10$^{-5}$(B) \\
\cline{2-3}
 & 63(a)$<$N$<$66(b) & IH with $\t>1.0\times10^{-2}$(G)\\
 &     & OR NH with $\t<1.1\times10^{-4}$(C) \\
\cline{2-3}
 & 66(b)$<$N$<$70(c) &
 NH with $7.8\times10^{-6}(A_2)<\t<2.5\times10^{-4}(F)$\\
 & & OR NH with $\t>1.3\times10^{-2}(H)$\\
\cline{2-3}
& N$>$70(c) & NH with $\t>2.5\times10^{-4}$(E) \\
\hline
 \end{tabular}
\caption
{\label{rate_stat}
The inferences on hierarchy and allowed range of $\theta_{13}$
that can be drawn from figure \ref{figrate_stat} from the observation
of total number of events.
We present the inferences for the panels 1,3 and 5 in figure
\ref{figrate_stat}. a, b, c, d, e correspond to the points shown in
figure \ref{figrate_stat}.
The bounds on $\theta_{13}$ correspond to vertical projections
from a, b, c, d, e on the $\theta_{13}$ axis from the points where
the horizontal lines meet the rate-curves.
}
\end{center}
\end{table}


The figure \ref{figrate_stat} 
plots the  total number of $\bar\nu_e+p$ and the $^{12}{C}$ 
CC  events observed in \kl as 
a function of $\tan^2\theta_{13}$ for the case with no earth matter effect 
(panels 1, 3 and 5),  
as well as for SN neutrinos passing through earths mantle and core 
(panels 2, 4 and 6). 
We also plot the statistical error bands (1$\sigma$) assuming 
$\sqrt{N}$ errors.  
These plots show that the
total number of charged current events can 
separate between
different mass hierarchies (at $1\sigma$)
for $\tan^2\theta_{13} > 2.3\times10^{-5}
$, $3.5\times10^{-5}$, $7.4\times10^{-5}$
for the $\bar\nu_e+ p$, $\bar\nu_e+ ^{12}{C}$
and $\nu_e +^{12}{C}$ reactions respectively  for the case of
no earth matter effect and for $\tan^2\theta_{13} > 2.9\times10^{-5}$,
$4.4\times10^{-5}$, $3.4\times10^{-5}$ for the case when we include 
earth matter effect. 
We notice that in 
the case of $\nue$ events, the difference between the number of 
events for NH and IH is enhanced with the introduction of 
earth matter effect 
if $\tan^2\theta_{13} \gsim$ a few times $10^{-5}$.
For neutrinos events with IH, 
$\peeih \sim \sin^2\theta_{12} \sim 0.3$ (cf. figure \ref{figpee13})
for the case of no earth effect whereas with earth effect it is 
$\peeih \sim P_{2e}^{\oplus} \sim (f_{reg} + \sin^2\theta_{12})$. 
We have checked that the factor $f_{reg}$ defined in 
Eq. (\ref{freg}) is mostly positive for all energies. 
This implies that \peeih 
is higher for neutrinos crossing the earth so that there is 
less flavour conversion  and 
hence from Eq. \ref{rcc} we see that 
the  number of events will be lower as seen in figure \ref{figrate_stat}.
For NH on the other hand, 
the earth regeneration effect is suppressed by the 
sharp fall of $P_H$ with $\theta_{13}$ and one does not observe a 
pronounced change in the number of events
due to earth effect in this range of 
$\tan^2\theta_{13}$. 
The net effect is an 
enhancement in the difference between the event rates 
for the two hierarchies in the case of $\nue$ events 
resulting in a better sensitivity 
towards separating between these.    
On the other hand, for $\anue$ events, the difference between 
the hierarchies is seen to reduce for neutrinos crossing the earth. 

To illustrate how the total number of events can be used to put bounds on 
$\theta_{13}$ and separate between the two  hierarchies, we draw  
horizontal projections on the number of events axes from the points at which 
the slope of the rate-curve changes. 
This is shown in figure \ref{figrate_stat} by the horizontal lines 
marked by small English alphabets a, b, c, etc. 
Next we drop 
 vertical projections on the 
$\tan^2\theta_{13}$ axis and 
find out the corresponding range of 
$\theta_{13}$.
The $\theta_{13}$ bounds are marked by the vertical projections 
A, B, C etc. 
In Table 
\ref{rate_stat} 
we give 
the observed number of events between the different horizontal 
lines and the inferences on hierarchy and $\theta_{13}$ that one can draw 
from these observations for the three CC reactions.
For purpose of illustration
we give this Table for the case when the neutrinos are not passing through 
the earth. However similar conclusions can also be drawn for the case when 
the time of occurrence of the SN 
makes the neutrinos traverse through the earth. 
We see from the Table that 
if we consider one particular reaction then  
the number of events lying between the different horizontal lines 
can:
 \\
(1)
identify the hierarchy uniquely and indicate the $\theta_{13}$ 
range, \\
(2)
allow both hierarchies and put some bound on $\theta_{13}$ for either or both 
of these. \\  
In the case when a unique solution is not obtained from one reaction 
it may help to take recourse to a second reaction channel.
For instance 
considering the 
$\anue+p$ reaction 
we find from Table \ref{rate_stat} that 
if the total event rate, $N < 485$  
then there are two possibilities:
\\
(i) the hierarchy is normal and no bound can be put on $\theta_{13}$,\\ 
(ii) the hierarchy is inverted and $\tan^2\theta_{13} 
\lsim 2.3 \times 10^{-3}$. 
\\
The ambiguity in hierarchy can however be resolved if
in the $\nue+^{12}{C}$ reaction one observes $N>66$. 
For this case 
one can definitely say that the hierarchy is normal 
and obtain a lower bound of $\tan^2\theta_{13} > 7.8 \times 10^{-6}$.
With the identification 
of the hierarchy as normal, 
the lower bound can be improved to $\tan^2\theta_{13} > 5.6 \times 10^{-6}$
if in the $\anue +^{12}{C}$ reaction one observes $29 -33$ events. 
Even when the number of events observed in one particular reaction 
gives a unique solution for the hierarchy and the  allowed  
range of $\theta_{13}$, it may be possible to use the  
observed number of events in another reaction effectively 
to further constrain the allowed $\theta_{13}$ range. 

We would like to point out here that this is an exploratory analysis 
which studies the sensitivity of the total number of events 
to $\theta_{13}$ and 
the mass hierarchy. 
The exact values of the bounds quoted in Table \ref{rate_stat} 
are not so important. 
An appropriate $\chi^2$ analysis procedure including all the 
uncertainties and parameter correlations  
will allow to study the 
conclusions obtained 
in more detail and 
with a greater
statistical significance.


\subsection{Using the charged current spectrum} 

\begin{table}
\begin{center}
\begin{tabular}{|c|c|c|c|}
\hline
energy & \multicolumn{3}{|c|}{Minimum detector volume (in kiloton)} \\
bins   &   \multicolumn{3}{|c|}{ required for  different reactions}\\
\cline{2-4}
in MeV & $\anue+p$ & $\anue+^{12}{C}$ CC & $\nue+^{12}{C}$ CC \\
\hline
21-26 & very large  & very large  & 8 \\
26-31 & 29 & very large & 11 \\
31-36 & 51 & 5 & 78 \\
36-41 & 10 & 88 & very large \\
41-46 & 2 & 9 & 5 \\
\hline
\end{tabular}
\caption{\label{minvol}
Minimum volume required for a scintillator detector
to statistically distinguish the number of events
between two  scenarios:
neutrinos undergoing no earth matter effects and
neutrinos passing through earth's core and mantle.}
\end{center}
\end{table}

\begin{figure}
\centerline{\psfig{figure=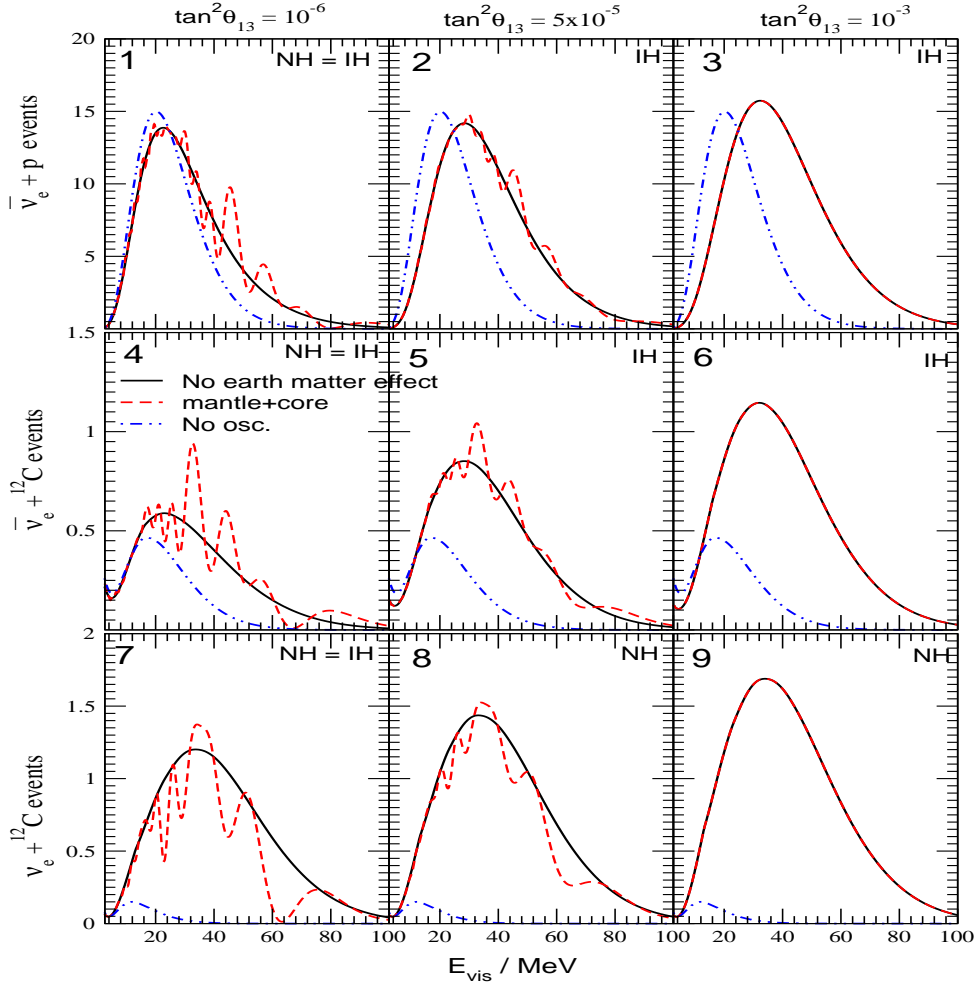,height=14cm,width=14cm,angle=0}}
\caption{
The CC spectra due to $\anue+p$, $\anue+^{12}{C}$ and $\nue+^{12}{C}$
events in KamLAND. The solid curves are for the case with
 no earth matter effect and the dashed curves are for 
the case when the neutrinos pass through 
earths mantle and core. The dotted lines correspond to 
the unoscillated spectrum.
We plot the spectrum for three sample values of 
$\theta_{13}$.
For the $\anue$($\nue$) events the spectra for NH(IH) are 
identical to the one presented for $\tan^2\theta_{13} 
= 10^{-6}$ and hence  these are not drawn for the higher 
$\theta_{13}$ values.} 
\label{spec}
\end{figure}

\begin{figure}
\centerline{\psfig{figure=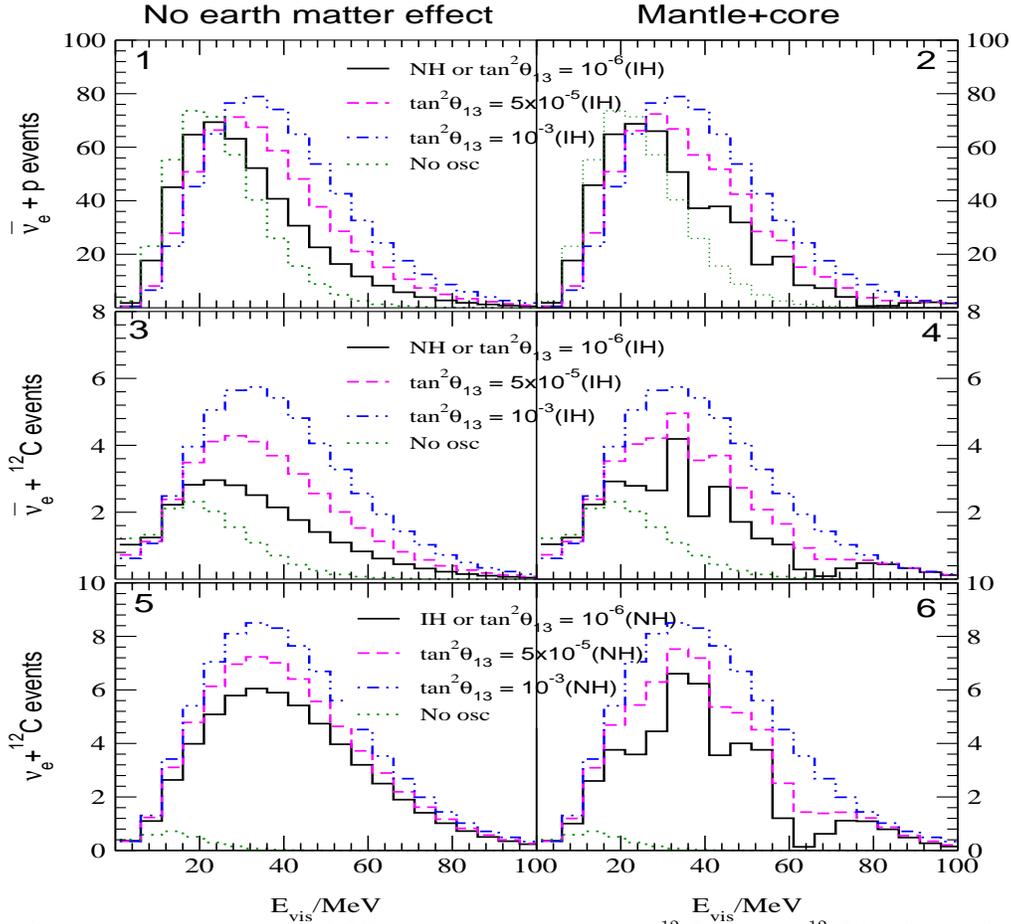,height=14cm,width=14cm,angle=0}}
\vskip -1cm
\caption{\label{figkl} Energy histogram plots of number of CC events
due to $\bar{\nu}_e+p$, $\bar{\nu}_e+^{12}C$ and $\nu_e+^{12}C$ 
in bins of width 5 MeV for the \kl detector.}
\end{figure}
 
\begin{figure}
\centerline{\psfig{figure=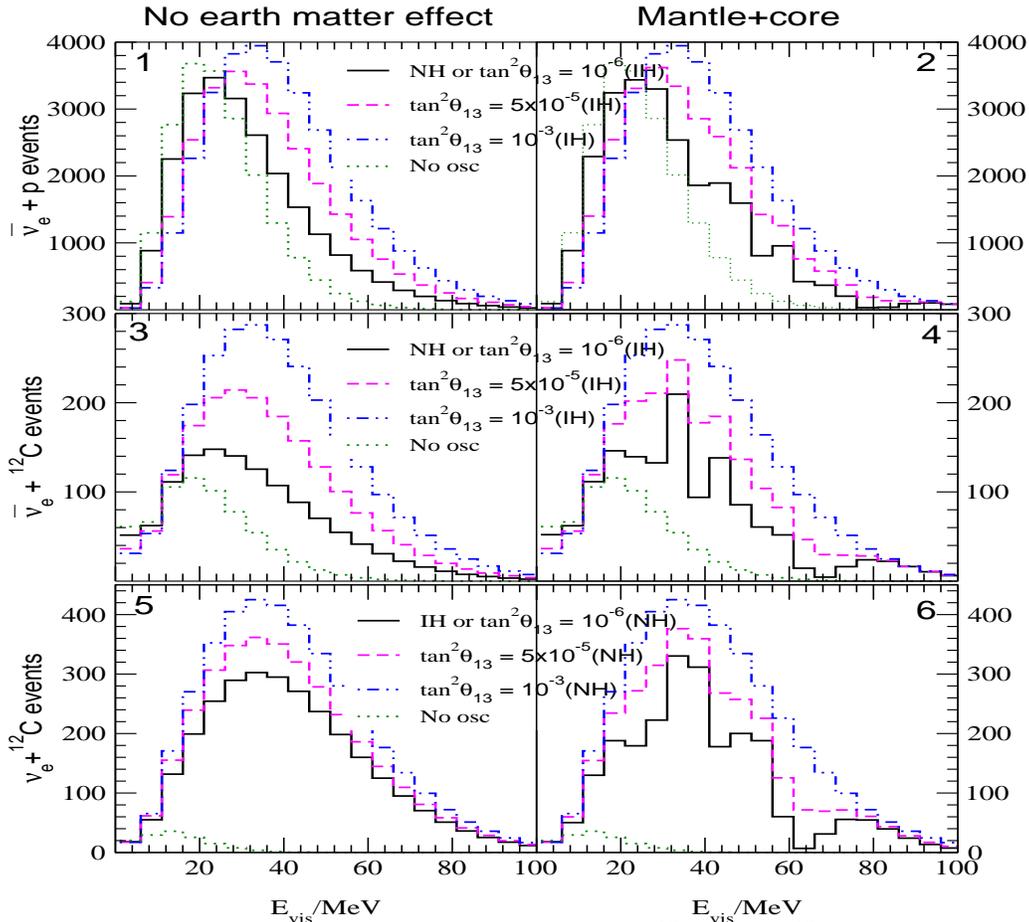,height=14cm,width=14cm,angle=0}}
\vskip -1cm
\caption{\label{lenafig} Energy histogram plots of number of CC events
due to $\bar{\nu}_e$+p, $\bar{\nu}_e+^{12}$C and $\nu_e+^{12}$C 
in bins of width 5 MeV for a 50 kilo ton scintillator detector
like LENA.}
\end{figure}


For the CC absorption of $\anue$ on free protons and 
$\anue$ and $\nue$ reactions on $^{12}{C}$, 
the detectors can observe the distribution
of the event rates with energy of the emitted electron or positron.
As the average energies of the 
$\nu_\mu$/$\nu_\tau$($\anumu$/$\anutau$)
are greater than the average energy of $\nu_e$($\anue$), 
the oscillated $\nu_e$($\anue$) will have a harder spectrum. 
This hardening of the energy spectrum of the neutrinos is manifested
in the observed positron/electron events in the detector, since
one has more high energy events compared to what would be
expected in absence of oscillations.
In addition, if the neutrino survival probabilities are energy 
dependent, then this would introduce a 
further distortion in the shape of the 
resultant spectrum. 
There are two possible sources of energy dependence in the survival 
probabilities: 
\\
(i) matter effect in the supernova -- this inducts an energy dependence in 
\peebarih and \peenh  through $P_H$,  
in the intermediate $\theta_{13}$ range; 
\\
(ii)matter effect inside the earth -- this leads to  
$\Delta m_{21}^2$ driven 
energy dependent frequency modulations.

In figure \ref{spec} we show the un-binned positron/electron 
spectrum for the ({\it i}) $ \anue+p$, ({\it ii}) $\anue+^{12}C$ and 
({\it iii}) $\nue+^{12}C$ reactions in KamLAND.
The dot-dashed curve gives the spectra for the case of no 
oscillation, the solid curves
are for the case where we have oscillation inside the SN but with
no earth matter effects, while the dashed
curves show the spectrum expected when the supernova neutrinos
cross the earth's mantle and core.
For the case where $\tan^2\theta_{13} = 10^{-6}$,
the  CC spectra for  
both $\nue$ and $\anue$ channel 
are identical for IH and NH (panels 1,4 and 7).  
For $\anue(\nue)$ flux with NH(IH) as the true hierarchy,
the event spectra
are not expected to change with 
$\theta_{13}$ in the range limited by the 
CHOOZ constraints. Therefore we present the  $\anue(\nue)$ spectra 
for NH(IH) only in the panels 1, 4 and 7. 
On the other hand the  $\anue(\nue)$ spectra 
for IH(NH) will depend on the value of $\theta_{13}$ 
both for with and without earth matter effects.
In the remaining panels we show the spectra for these 
cases for two different
values of $\tan^2\theta_{13}$ -- $5\times 10^{-5}$ and $10^{-3}$. 
We also note from this figure the 
fast variation of the spectrum 
shape due to earth matter effects. 

In figure \ref{figkl}
we show the expected energy spectrum 
for the three possible CC reactions
in \kl in bins of 5 MeV width. The left-hand panels show the 
expected spectra without earth matter effect while the 
right-hand panels display what we expect in case the neutrino cross 
the earth.
For both the $\anue$ and $\nue$ events, the solid curve denotes the case of
$\tan^2\theta_{13} = 10^{-6}$ and it corresponds to both NH and IH.
For $\anue(\nue)$ events with NH(IH) the spectra remains the same 
as that shown by the solid lines for all values of $\theta_{13}$, 
however for IH(NH) the spectra will change with $\theta_{13}$.
These are shown by the dashed and dot-dashed lines for 
two different values of $\tan^2\theta_{13}$.
The dotted lines show the no-oscillation spectra.

\subsubsection{No earth matter effect}
 
The comparison of the curves corresponding to the case of no 
oscillation in figures \ref{spec} and \ref{figkl} 
to the ones with oscillations, show the effect of oscillations 
in general on the expected spectrum in KamLAND.
%

For a given hierarchy, say IH for $\anue$ (panels 1 and 3 of 
figure \ref{figkl}) or
NH for $\nue$ (panel 5 of figure \ref{figkl}), 
if we compare the spectra 
for different values of $\tan^2\theta_{13}$,
we note an increase in the number of events in 
higher energy bins, with the increase in the value of 
$\tan^2\theta_{13}$.
This is due to the change in the
relevant survival probability due to matter effect in the SN.
For $\tan^2\theta_{13}\sim 10^{-6}$, $P_H=1$ 
and we have only the lower
resonant conversion in the SN. As $\tan^2\theta_{13}$ increases,
$P_H$ decreases and eventually becomes zero at
$\tan^2\theta_{13}\sim 10^{-3}$.
A small $P_H$ implies a lower survival probability,
more flavor conversion and hence harder spectrum for the
incoming (anti)neutrinos. This
change in the shape of the spectrum with $\theta_{13}$ can be
used to limit the value of $\tan^2\theta_{13}$ for
a given hierarchy.
Also, since for NH(IH), the $\anue$($\nue$) spectrum are independent
of $\tan^2\theta_{13}$ and correspond to the 
the curves shown by bold solid lines in figure \ref{figkl},
a comparison of the observed spectrum
against the expected ones
can shed some light on the neutrino mass hierarchy, if
true value of $\tan^2\theta_{13}$
is $\gsim 10^{-6}$.
Note however that the value of
$\tan^2\theta_{13}$ and the neutrino mass hierarchy come
correlated. For instance, if the observed spectrum corresponds to
the one shown by the solid black line in panel 1 of figure \ref{figkl}
for $\anue+p$ events,
then it would be difficult to decide whether the hierarchy
is inverted and $\tan^2\theta_{13} \lsim 10^{-6}$  or the hierarchy 
is normal with no bound on $\theta_{13}$, using only this reaction.
However, if in addition
the observed spectral shape corresponding to
electron events generated by the $\nue + ^{12}C$ reaction 
conforms to the one shown by, say, the dot-dashed line in 
panel 5 of figure \ref{figkl},
then we can say that
the mass hierarchy is normal and one can put a lower
bound on $\tan^2\theta_{13}$.

\subsubsection{Including earth matter effect}

Matter effects inside the earth produce
$\Delta m_{21}^2$ driven
oscillations
\cite{Dighe:1999bi,Lunardini:2001pb,Lunardini:2003eh,
Dighe:2003jg,Dighe:2003vm}
which we see in the dashed curves shown in figure \ref{spec} 
and right-hand panels in figure \ref{figkl}.
These high frequency oscillations in the energy spectrum are
superimposed over the dominant spectral distortion due to transitions
in the supernova.
Observation of these fast modulations in the observed spectrum would
give an unambiguous signature of the presence of earth effects,
which can be used to constrain the hierarchy and $\theta_{13}$.
In particular, we note that there are no earth
induced wiggles in the resultant $\anue(\nue)$ spectrum for
$\tan^2\theta_{13} \gsim 10^{-3}$ for IH(NH). However, if the
hierarchy is normal(inverted) then even if $\tan^2\theta_{13} \gsim
10^{-3}$,
one would expect earth matter effects in the $\anue(\nue)$ spectrum
as is evident from panels 1, 4 and 7 of figure \ref{spec} and 
the solid black lines in figure \ref{figkl},
which show the spectra for these cases. Thus the
detection of earth matter effects in the
antineutrino(neutrino) channel for $\tan^2\theta_{13} \gsim 10^{-3}$
would clearly
indicate normal(inverted) hierarchy for the neutrino mass spectrum.
If we use the earth matter signatures from
just the neutrino or the antineutrino channel alone,
we would need some prior information that $\tan^2\theta_{13} \gsim 10^{-3}$.
This could come either from the total rates signature of the
supernova signal itself (as discussed in the previous sub-section)
or from some other terrestrial experiment.
However, if we use the neutrino and antineutrino channel simultaneously
and look for earth matter effects, we could pin down both the
neutrino mass hierarchy and the range of
$\tan^2\theta_{13} \gsim 10^{-3}$, if this indeed was the true
range of $\tan^2\theta_{13}$.
For example,
if we observe no earth matter dependent modulations in the
neutrino channel and at the same time observe the earth
matter generated wiggles in the antineutrino channel, we could
surely say that the hierarchy is normal and
$\tan^2\theta_{13} \gsim 10^{-3}$.
 
\subsubsection{Detection of earth matter effect induced modulations}

The figure \ref{spec}
reflects the fact that the earth effect induced
modulations are more rapid but of less amplitude
in the low energy regime, whereas in the high  energy
domain they are less rapid but of greater amplitude.
Since in the low energy regime the
modulations are very frequent in energy, they are averaged
out even in a very narrow energy bin. So it is not possible for any
detector to observe the ``earth matter effect induced
modulations'' in  the lower energy bins (cf. figure \ref{figkl}).
On the other hand, towards the higher energy regime,
the modulations are highly peaked and more wide
(roughly the same order as the width of the energy bins we take),
and hence may be  observed in the event distribution,
as can be seen in figure \ref{figkl}.
If the energy resolution is poor
which means a measured electron/positron energy
corresponds to a wide range of true energy (and hence
neutrino energy) then even the wide modulations
at high energy domain
may be averaged out within this range and consequently
these modulations can't be seen even in the higher energy bins.
So a detector needs a very good energy resolution
to detect the earth matter effect induced modulations.
\kl being a scintillation detector
fulfills this criterion very well and we note from 
the right-hand panels of figure \ref{figkl} that 
\kl does see the earth induced modulations in the energy spectrum 
at the high energy end. 
However, even though the event rate in \kl 
for the $\anue+p$ reaction is not very low, the 
statistics in \kl in the individual spectral bins 
for the $^{12}C$ reactions are poor with the present volume of 1 kton. 
This could therefore preclude
any statistically significant signature for earth matter effects in
KamLAND at least in the $^{12}{C}$ reaction channels. 
We therefore need larger \kl like scintillation detectors for
observing the earth matter effects in the detected supernova
spectrum, where 
these $^{12}C$ reactions could be probed to 
look for earth matter effects.
In Table \ref{minvol} we give the minimum
detector mass required for a \kl
type scintillator detector to
observe statistically
significant earth matter effect induced
wiggles in various energy bins.
The Table \ref{minvol} shows that in general
we need lower detector mass to detect the
modulations in the higher energy bins.
For the lower energy
bins (which have a better statistics)
the $\nu_e+^{12}C$ reaction is most promising  and can detect the
earth matter effect modulations with a much smaller volume detector.

\subsubsection{The CC spectrum for a 50 kton detector}

Even though from figure \ref{figkl} we see that the statistics 
for the $\anue+p$ reaction is not very poor in KamLAND, this reaction 
feels lesser impact of the earth generated modulations of the 
spectrum in the energy bins near the peak.
On the other hand the $^{12}C$ reactions have greater fluctuations
over the no earth effect spectra in the energy bins near the peak. 
Thus the $^{12}C$ 
reactions seem to have a better potential to
probe earth effects in the resultant spectrum.
But one needs a larger detector to study the 
energy spectra of these reactions and make a statistically 
significant statement. 
In  figure \ref{lenafig} we show the corresponding 
expected energy binned spectrum in a 50 kton
liquid scintillation detector.
Recently a large liquid scintillation detector,
LENA, has been proposed which would have a total mass of 50 kton and
a fiducial mass of about 30 kton \cite{lena}. 
We use the largest possible detector mass for the sake of illustration. 
This figure is a ``scaled up'' version of figure \ref{figkl} and 
all the features discussed in the context of figure \ref{figkl}
appear in a much magnified form here. In particular,
the fast oscillations imprinted on the spectra $^{12}C$ 
corresponding to $\anue$ and $\nue$ reactions can be combined 
to lead to unambiguous determination of the hierarchy and 
limit on the value of $\theta_{13}$, as discussed before for 
KamLAND. For LENA it  
can be verified with a good statistical precision.
The $\anue+p$ process in LENA would have a statistics comparable 
to that expected in SK --  with much better energy resolution and 
hence greater spectral power.


\section{SN uncertainties and CC/NC ratios}

\begin{figure}
\centerline{\psfig{figure=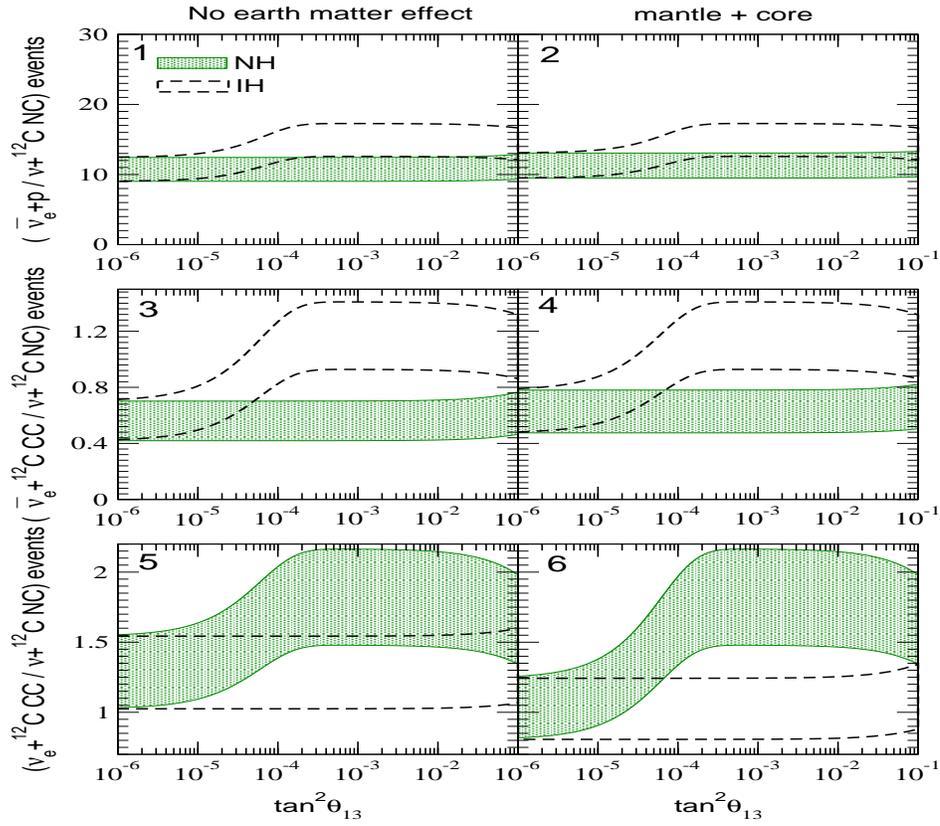,height=12cm,width=14cm}}
\caption{Plot of ratio of $\nu_e$($\bar{\nu}_e$)+
$^{12}C$ CC event rates to ($\nu$+$^{12}C$) NC event rates
 as a function of
$\theta_{13}$ for NH and IH, including the statistical
uncertainty}
\label{figratio_stat}
\end{figure}
 
\begin{figure}
\centerline{\psfig{figure=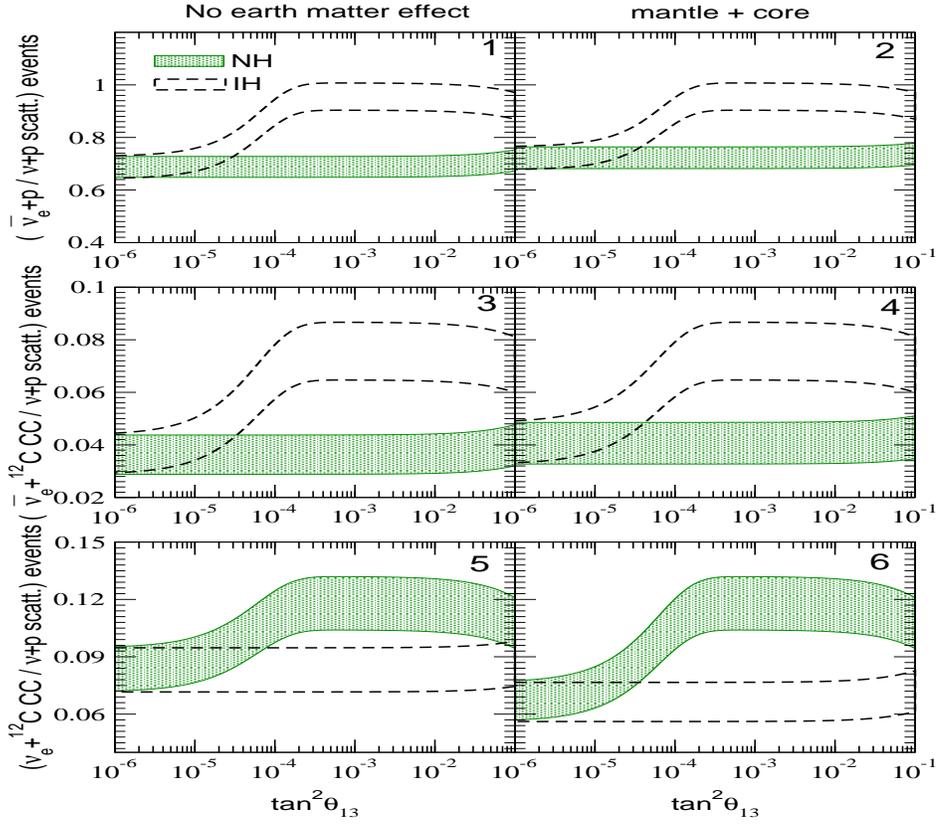,height=12cm,width=14cm,angle=0}}
\caption{Plot of ratio of $\nu_e$($\bar{\nu}_e$)+
$^{12}C$ CC event rates to ($\nu+p$) scattering event rates
 as a function of
$\theta_{13}$ for NH and IH, with the
statistical uncertainty.}
\label{figrationup_stat}
\end{figure}

In the previous sections we have considered the effect of statistical
errors only. However there are uncertainties emerging from our (lack of)
knowledge of
neutrino fluxes in a supernova 
due to: 
\\
$\bullet$ uncertainty in the total value of neutrino luminosity \\
$\bullet$ uncertainty in the luminosity distribution between the 6 species \\
$\bullet$  uncertainty in the average temperature of each neutrino
species\\
$\bullet$  uncertainty in the value of pinching factor in the
neutrino spectra which leads to deviation from the
Fermi-Dirac distribution \\
In our calculation so far we have used 
$L_e = L_{\bar{e}} = L_x = L_{\bar{x}}$ where $x$ can be $\mu$ or $\tau$.
However these can vary as $L_{e/\bar e} = (0.5 -2 ) L_{x/\bar x}$ 
with $L_{x}$ and $L_{\bar{x}}$
being equal \cite{Keil:2002in}. 
The average energies of $\nu_e$, $\anue$ and $\nu_x$ can vary in the range 
($7 -18$) MeV, ($14-22$) MeV and ($15 - 35$) MeV \cite{Keil:2002in}. 
The pinching factor can take values between ($0-2$) for $\nu_x$ 
and ($0-3$) for $\nu_e$ and $\anue$. 
Finally the absolute luminosity can be uncertain
in the range  $(1-5)\times 10^{52}$ 
ergs. 
Incorporation of these uncertainties can change the above conclusions. 
These uncertainties are correlated and a proper
treatment requires a $\chi^2$ analysis incorporating these 
uncertainties in a correlated fashion. 
In this section we explore if it is possible to construct variables which 
are free from some of the uncertainties and introduce the  
CC/NC ratios. We study the capability of these ratios 
in separating between the normal and inverted hierarchies 
and also their sensitivity to $\theta_{13}$. 
These ratios have the advantage that the uncertainty due to the 
total luminosity 
gets canceled  out from the numerator and the 
denominator. 

As we have discussed in section 4, \kl is sensitive to two NC reactions
(i) $\nu+^{12}C$ and (ii) $\nu+p$. 
In figure 
figure \ref{figratio_stat} we plot the 
CC/NC ratios
as a function of $\theta_{13}$, for normal and inverted hierarchy with
1$\sigma$ statistical  error bands.
The NC reaction considered is the $\nu+^{12}C$ reaction.
In figure \ref{figrationup_stat}  
we plot the same thing but using the $\nu-p$ scattering reaction 
as the NC reaction channel.


\begin{table}
\begin{center}
\begin{tabular}{|c|c|c|c|}
\hline
Reactions & CC events  & CC/$\nu+^{12}C$ NC events & CC/$\nu$+p scatt. events \\
\hline
$\bar{\nu}_e+p$ & 3.94(4.66) & 15.72(15.92) & 5.51(6.04)\\
$\bar{\nu}_e+^{12}C$& 14.11(20.31) &20.75(25.38) & 16.22(22.94) \\
$\nu_e+^{12}C$ & 13.42(11.12) & 20.39(18.90) & 14.33(11.96)\\
\hline
\end{tabular}
\caption{\label{table_error}The 1$\sigma$ 
\% spreads defined in  Eq. (\ref{spread}) 
at  $\tan^2\theta_{13}=10^{-3}$ for IH(NH).}
\end{center}
\end{table}
\begin{table}
\begin{center}
\begin{tabular}{|c|c|c|c|}
\hline
Reactions & CC events  & CC/$\nu+^{12}C$ NC events & 
CC/$\nu+p$ scatt. events \\
\hline
$\bar{\nu}_e+p$ & 11.81 & 0 & 10.51\\
$\bar{\nu}_e+^{12}C$& 19.06 & 13.32 & 18.91\\
$\nu_e+^{12}C$ & 4.73 & -1.63 & 4.50\\
\hline
\end{tabular}
\caption{\label{table_separation}The ``separation''
as defined in Eq. (\ref{sep}) for different CC events
as well as for the CC/NC ratios.
The separations in this Table are calculated at a sample value 
of $\tan^2\theta_{13} = 10^{-3}$.}
\end{center}
\end{table}


It is instructive to look at the \%-spread in the widths of the 
error bands relative to the central value for each case as 
well as for the CC rates in figure \ref{figrate_stat}.  
The \% spread is defined as 
\beq
{\rm spread} = \frac{R_{upper} - R_{lower}}
{\rm R_{central}}\times 100
\label{spread}
\eeq
where $R$ denotes the values of the event rates or ratios 
at the appropriate limit 
denoted by the suffix. 
When we take the ratio of CC event rates(x) to the NC event rate(y),
the \% spread is  
$\sqrt{(1/x+1/y)}$. 
Therefore  larger the NC event rate smaller is the spread and so 
the $\nu-p$ reactions are expected to have smaller spread  
than the $\nu-^{12}{C}$ reaction. 
It is relevant to point out here that the \% spread for the 
the CC rates goes as $\sqrt{(1/x)}$ and hence it will have a narrower 
\% spread as far as statistical error is concerned. 
In Table \ref{table_error} we show the \% spreads for $\tan^2\theta_{13}
= 10^{-3}$ for the three reactions. 
The Table shows that
the \% spread is least for 
the CC rates. 
However the effect of the absolute luminosity uncertainty 
(from which the ratios are free)  
may offset this advantage for the CC rates.   

When we study the effectiveness with which one can separate between 
normal and inverted hierarchies it is useful to define the 
quantity 
\beq
\rm separation = 100\times\left(\frac{R^{H_1}_{lower}-R^{H_2}_{upper}}
{R^{IH}_{central}+R^{NH}_{central}}\right)\\
\label{sep}
\eeq
where for $\anue$($\nue$) events the suffixes $H_1$ and $H_2$ in the 
first  and the second term in the numerator will 
correspond  to IH(NH) and NH(IH) respectively.  
The suffix ``central'' denote the 
the central values of event rates or ratios.
The subscripts ``upper'' and ``lower'' are used to indicate respectively
the (central + 1$\sigma$)  and (central - 1$\sigma$) values.

In Table \ref{table_separation} we show this quantity for the CC rates as 
well as for the ratios for purposes of 
illustration for the specific case of  $\tan^2\theta_{13} = 10^{-3}$ and 
with no earth matter effect.  
It is clear from the table that at least at this value of $\theta_{13}$, 
the CC/NC ratios w.r.t the $\nu-p$ NC scattering does an even better job 
than the CC rates, with the added 
advantage that the ratios are free from the 
uncertainties due to absolute  luminosity. 
From figures \ref{figrationup_stat} (and also figure \ref{figrate_stat})
we can convince ourselves that
this conclusion is valid for other values of $\theta_{13}$ as well, 
if $\theta_{13}$ is not too small, for which in any case the  
SN neutrinos loose the  sensitivity to separate between the hierarchies. 
As we have seen earlier with the introduction of earth matter 
effect for the $\nue$ events there is more earth regeneration for 
IH implying there is less conversion and the number of events for IH 
decreases and hence the     
difference between IH and NH increases  
(cf. fig. 2) as compared to the no earth matter effect case. 
Thus when neutrinos pass through earth, the $\nue$ events gain
a better capability of distinguishing between the hierarchies. 

The Table \ref{table_separation} 
and the figure \ref{figratio_stat} 
show that 
when the ratio is taken w.r.t the $\nu-^{12}{C}$ reaction 
the 
potential
of the the CC/NC variable to distinguish between normal and inverted
hierarchies worsens for the $\anue+p$ and the $\nue+^{12}{C}$ reaction 
and the NH and IH becomes indistinguishable excepting the case  
of panel 6 in figure \ref{figratio_stat} 
for the $\nue +^{12}{C}$ reaction where earth matter effect helps. 


\section{Summary and Conclusions}

Although the neutrino fluxes from supernova suffer from inherent astrophysical 
uncertainties which can be large, the physics potential offered 
by them are quite rich. A core-collapse supernova provides 
an unique environment where one can study the propagation and matter 
effect of all three species of neutrinos and antineutrinos. 
The varying density in supernova allows the occurrence of two MSW 
resonances corresponding to both atmospheric and solar mass scale and
if a detector is placed such that the neutrinos 
from the supernova traverse through the earth's mantle and/or core
then earth matter effects also come into play.
 
As far as the propagation in SN matter is concerned, 
the solar neutrino oscillation parameters lying in the LMA region
govern the lower resonance and the propagation is fully adiabatic.
Propagation through the higher 
resonance could be non-adiabatic depending on the hierarchy and  
$\theta_{13}$. 
For $\anue(\nue)$ with IH(NH) 
there are three possibilities determined by the 
mixing angle $\theta_{13}$:
\\
(i) If $\tan^2\theta_{13} \lsim 10^{-6}$, the propagation at higher resonance 
is completely non-adiabatic, $P_H=1$, 
and the survival probability is identical for IH and NH 
for both $\nue$ and $\anue$.
In this regime it is not possible to differentiate between NH and IH 
and only an upper limit on $\theta_{13}$ can be given. 
\\
(ii) For $ 10^{-6} \lsim \tan^2\theta_{13} \lsim 10^{-3}$, 
the propagation is partially non-adiabatic for neutrinos
(antineutrinos) and normal(inverted) hierarchy 
and depends on $P_H$. 
$P_H$ changes from 1 to 0 depending 
on the value of $\theta_{13}$ and
one can  
constrain $\theta_{13}$ within a range. 
\\
(iii) For $\tan^2\theta_{13} \gsim 10^{-3}$ the propagation is completely 
adiabatic($P_H=0$) for the $\nue$($\anue$) and normal(inverted) hierarchy 
and flavour conversion is maximum.  
The  probabilities do not vary with $\theta_{13}$ in the 
range permitted by the CHOOZ data. 
In this regime one can therefore set only a lower bound on $\theta_{13}$.  
\\
(iv) 
The $\nue$($\anue$)
propagation in SN matter for inverted(normal) hierarchy  
have no dependence 
on $P_H$ and the survival probabilities 
do not change with  
$\theta_{13}$. 
Therefore if $\theta_{13}$ 
is such that  the value of 
$P_H$ is
expected to be 
different from 1 
for $\nue$($\anue$) and normal(inverted)   
then observation of total number of events can 
distinguish between the hierarchies in both neutrino and antineutrino channel. 
The power of discrimination between the two hierarchies is maximum in the 
region (iii) above where $P_H$ differs maximally from 1. 
  
This dependence of the probabilities on $\theta_{13}$ and hierarchy 
gives SN neutrinos (at least in principle) remarkable sensitivity  
to probe very small values of $\theta_{13}$, much smaller than the 
projected limits from the reactor experiments,
superbeams and even neutrino factories, 
and also the 
mass  hierarchy. 

We study in detail the sensitivity of the 
the total observed event rates, in the $\anue+p$, $\anue+^{12}{C}$ 
and $\nue+^{12}{C}$ reactions in the \kl detector 
to probe $\theta_{13}$ and to distinguish between 
normal and inverted hierarchy. 
We find that considering only the statistical errors into account, 
it may be possible to uniquely determine 
the hierarchy and give a bound on the allowed range of $\theta_{13}$.
In some cases it may not be possible to disentangle the correlated 
information on hierarchy and the allowed range of $\theta_{13}$ 
from the observation of total number of events in one reaction channel. 
However a distinction may be possible 
using the observed event rates in another reaction. 
We find that it is particularly useful to combine the 
information from $\anue$ and $\nue$ reaction channels. 

Apart from the total event rates it is also possible 
to observe 
the CC spectrum due to (i) $\anue+p$, (ii) $\anue +^{12}{C}$ and 
(iii) $\nue+^{12}{C}$ 
reactions in KamLAND. 
Since the average energies of muon and tau type neutrinos 
are more, the flavor 
converted electron type neutrinos and antineutrinos 
will have a harder spectrum even if the probabilities 
are energy independent. 
The matter effect in 
the SN may induce an additional energy dependence in the probabilities 
$\peenh$ and $\peebarih$ 
if $\theta_{13}$ lies in the intermediate range. 
As $\theta_{13}$ increases there is more flavor conversion 
giving rise to more neutrinos in the higher energy bins.  
This hardening of the spectrum can be used to put bounds on $\theta_{13}$. 
On the other hand 
$\nu_e$ events for IH and $\anue$ events for NH 
do not display this extra $\theta_{13}$ 
dependent hardening of the spectrum.
This can therefore be used to 
separate between the hierarchies. 

If in addition there is earth matter effect
then it  
gives rise to fast
modulations of the event rate with energy. 
For the $\anue(\nue)$ spectra, while 
these modulations are independent of $\theta_{13}$ for NH(IH),
for IH(NH) they are decrease with the rise of 
$\theta_{13}$ and vanish for $\tan^2\theta_{13} \gsim 10^{-3}$.
Therefore observation of these earth effect induced modulations
give a handle to probe $\theta_{13}$ and the true mass hierarchy. 

To observe this modulations one needs a very good energy resolution
and \kl being a scintillator detector is ideal for the purpose. 
We also study the potential of the capture reactions of $\nue$ and $\anue$ 
on $^{12}C$,  for observing the earth matter induced 
fast fluctuations in the observed spectrum.
However these reactions have low statistics in \kl which 
has only 1 kton of detector mass.
We determine the  minimum volume that is required to observe
the modulations due to earth matter effect in a generic 
scintillation detector in various energy bins. 
We present for a planned 50 kton detector the energy spectrum for various 
values of $\theta_{13}$ and discuss how the energy 
spectrum can be used to differentiate 
between $\theta_{13}$ and also to discriminate between the hierarchies. 

We also construct the CC/NC ratios which are better as far as the 
SN uncertainties are concerned, 
since the absolute uncertainties can get
canceled in these ratios. 
However since the ratios have a larger relative statistical error 
their usefulness is limited unless the statistics of the NC events are very
large. The two NC reactions that the SN neutrinos can undergo in \kl 
are $\nu-^{12}C$ and $\nu-p$ reaction.
Since the latter has a far better statistics, 
it is superior to the $\nu-^{12}C$ reaction for probing $\theta_{13}$ 
and hierarchy. 
We find that the CC/NC ratio w.r.t the $\nu-p$ scattering reaction 
even has a better power to separate between the hierarchies  
compared 
to the only CC rates.

To conclude, in this paper we have made an in-depth  study of the 
potential of the \kl detector in probing $\theta_{13}$ and sign of the 
atmospheric mass scale $\Delta m^2_{32}$ through observation of SN 
neutrinos. 
The results look promising 
and warrants a more rigorous statistical analysis simulating the 
supernova signal and including the astrophysical 
uncertainties with their proper correlation.  

\vskip 0.5cm
\centerline{\Large\bf{Appendix A}}
\begin{figure}
\centerline{\psfig{figure=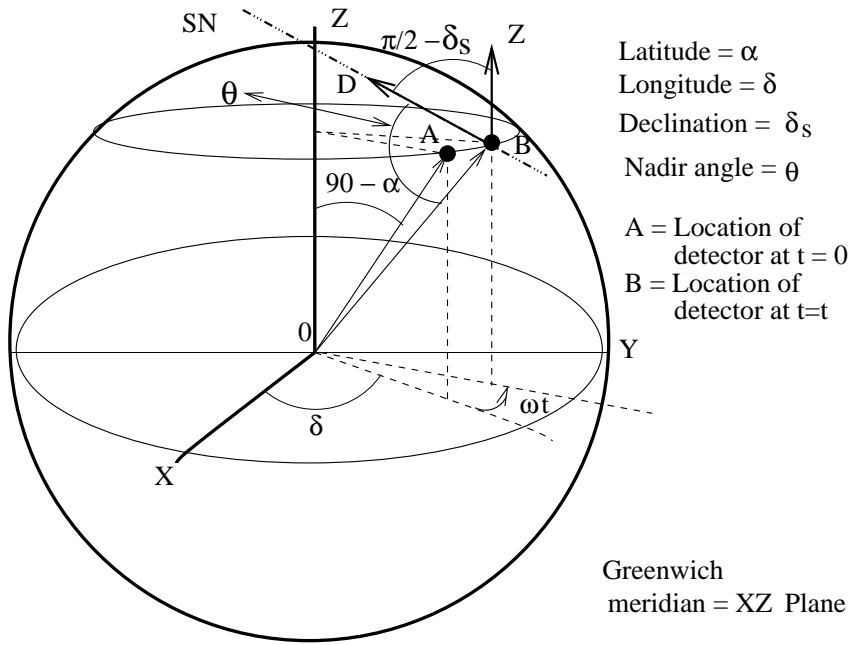,width=10cm,angle=270}}
\caption{The geometry describing the trajectory of a neutrino
from supernova inside the earth in reaching a detector. }
\label{figearth}
\end{figure}

For a given location of the supernova the nadir angle
subtended at a particular detector also changes continually with time
as the earth precesses about it axis of rotation with an angular speed
of $\frac{2\pi}{24}$ hours$^{-1}$. The figure \ref{figearth}
describes this geometry. The direction of the supernova
from earth can be specified by the angle $\delta_s$ (the declination),
which is the angle that the supernova direction makes with the plane
of the equator. We denote the latitude and longitude of the detector
by ($\alpha$,$\delta$) and the angular speed of earth's rotation
about its own axis by $\omega$. We define an universal time which is
counted from a moment (t=0) when the supernova is on the
Greenwitch meridian.Then one can write the nadir angle $\theta_n$
of the detector at time t as
\bea
\cos\theta_n &=&   \label{nadir}
 - ( \cos\alpha \cos\delta_s \cos(\omega t + \delta)
  + \sin\alpha \sin\delta_s )
\eea

For $\cos\theta_n < 0$ the SN neutrinos do not cross the earth,
for  $0< \cos\theta_n< 0.84$ the neutrinos pass through 
the earth's mantle, while for
$0.84<\cos\theta_n<1$ they pass through
both the mantle and core of the earth.

\vskip 10pt

S.G. would like to thank Mark Vagins for useful correspondences.  
S.C. acknowledges discussions with Lothar Oberauer on the proposed 
LENA detector.


\end{document}